# Studying the Underlying Event in Drell-Yan and High Transverse Momentum Jet Production at the Tevatron

T. Aaltonen,[26] J. Adelman,[65] B. Álvarez González[v,13] S. Amerio[dd,46] D. Amidei,[37] A. Anastassov,[41] A. Annovi,[22] J. Antos,[17] G. Apollinari,[20] A. Apresyan,[51] T. Arisawa,[61] A. Artikov,[18] J. Asaadi,[57] W. Ashmanskas,[20] A. Attal,[4] A. Aurisano,[57] F. Azfar,[45] W. Badgett,[20] A. Barbaro-Galtieri,[31] V.E. Barnes,[51] B.A. Barnett,[28] P. Barria[ff,49] P. Bartos,[17] G. Bauer,[35] P.-H. Beauchemin,[36] F. Bedeschi,[49] D. Beecher,[33] S. Behari,[28] G. Bellettini[ee,49] J. Bellinger,[63] D. Benjamin,[19] A. Beretvas,[20] D. Berge,[16] A. Bhatti,[53] M. Binkley,[20] D. Bisello[dd,46] I. Bizjak[jj,33] R.E. Blair,[2] C. Blocker,[7] B. Blumenfeld,[28] A. Bocci,[19] A. Bodek,[52] V. Boisvert,[52] D. Bortoletto,[51] J. Boudreau,[50] A. Boveia,[66] B. Brau[a,66] A. Bridgeman,[27] L. Brigliadori[cc,6] C. Bromberg,[38] E. Brubaker,[65] J. Budagov,[18] H.S. Budd,[52] S. Budd,[27] K. Burkett,[20] G. Busetto[dd,46] P. Bussey,[24] A. Buzatu,[36] K. L. Byrum,[2] S. Cabrera[x,19] C. Calancha,[34] S. Camarda,[4] M. Campanelli,[33] M. Campbell,[37] F. Canelli[14,20] A. Canepa,[48] B. Carls,[27] D. Carlsmith,[63] R. Carosi,[49] S. Carrillo[n,21] S. Carron,[20] B. Casal,[13] M. Casarsa,[20] A. Castro[cc,6] P. Catastini[ff,49] D. Cauz,[58] V. Cavaliere[ff,49] M. Cavalli-Sforza,[4] A. Cerri,[31] L. Cerrito[q,33] S.H. Chang,[30] Y.C. Chen,[1] M. Chertok,[8] G. Chiarelli,[49] G. Chlachidze,[20] F. Chlebana,[20] K. Cho,[30] D. Chokheli,[18] J.P. Chou,[25] K. Chung[o,20] W.H. Chung,[63] Y.S. Chung,[52] T. Chwalek,[29] C.I. Ciobanu,[47] M.A. Ciocci[ff,49] A. Clark,[23] D. Clark,[7] G. Compostella,[46] M.E. Convery,[20] J. Conway,[8] M.Corbo,[47] M. Cordelli,[22] C.A. Cox,[8] D.J. Cox,[8] F. Crescioli[ee,49] C. Cuenca Almenar,[64] J. Cuevas[v,13] R. Culbertson,[20] J.C. Cully,[37] D. Dagenhart,[20] M. Datta,[20] T. Davies,[24] P. de Barbaro,[52] S. De Cecco,[54] A. Deisher,[31] G. De Lorenzo,[4] M. Dell'Orso[ee,49] C. Deluca,[4] L. Demortier,[53] J. Deng[f,19] M. Deninno,[6] M. d'Errico[dd,46] A. Di Canto[ee,49] G.P. di Giovanni,[47] B. Di Ruzza,[49] J.R. Dittmann,[5] M. D'Onofrio,[4] S. Donati[ee,49] P. Dong,[20] T. Dorigo,[46] S. Dube,[55] K. Ebina,[61] A. Elagin,[57] R. Erbacher,[8] D. Errede,[27] S. Errede,[27] N. Ershaidat[bb,47] R. Eusebi,[57] H.C. Fang,[31] S. Farrington,[45] W.T. Fedorko,[65] R.G. Feild,[64] M. Feindt,[29] J.P. Fernandez,[34] C. Ferrazza[gg,49] R. Field,[21] G. Flanagan[s,51] R. Forrest,[8] M.J. Frank,[5] M. Franklin,[25] J.C. Freeman,[20] I. Furic,[21] M. Gallinaro,[53] J. Galyardt,[14] F. Garberson,[66] J.E. Garcia,[23] A.F. Garfinkel,[51] P. Garosi[ff,49] H. Gerberich,[27] D. Gerdes,[37] A. Gessler,[29] S. Giagu[hh,54] V. Giakoumopoulou,[3] P. Giannetti,[49] K. Gibson,[50] J.L. Gimmell,[52] C.M. Ginsburg,[20] N. Giokaris,[3] M. Giordani[ii,58] P. Giromini,[22] M. Giunta,[49] G. Giurgiu,[28] V. Glagolev,[18] D. Glenzinski,[20] M. Gold,[40] N. Goldschmidt,[21] A. Golossanov,[20] G. Gomez,[13] G. Gomez-Ceballos,[35] M. Goncharov,[35] O. González,[34] I. Gorelov,[40] A.T. Goshaw,[19] K. Goulianos,[53] A. Gresele[dd,46] S. Grinstein,[4] C. Grosso-Pilcher,[65] R.C. Group,[20] U. Grundler,[27] J. Guimaraes da Costa,[25] Z. Gunay-Unalan,[38] C. Haber,[31] S.R. Hahn,[20] E. Halkiadakis,[55] B.-Y. Han,[52] J.Y. Han,[52] F. Happacher,[22] K. Hara,[59] D. Hare,[55] M. Hare,[60] R.F. Harr,[62] M. Hartz,[50] K. Hatakeyama,[5] C. Hays,[45] M. Heck,[29] J. Heinrich,[48] M. Herndon,[63] J. Heuser,[29] S. Hewamanage,[5] M. Hickman,[12] D. Hidas,[55] C.S. Hill[c,66] D. Hirschbuehl,[29] A. Hocker,[20] S. Hou,[1] M. Houlden,[32] S.-C. Hsu,[31] R.E. Hughes,[42] M. Hurwitz,[65] U. Husemann,[64] M. Hussein,[38] J. Huston,[38] J. Incandela,[66] G. Introzzi,[49] M. Iori[hh,54] A. Ivanov[p,8] E. James,[20] D. Jang,[14] B. Jayatilaka,[19] E.J. Jeon,[30] M.K. Jha,[6] S. Jindariani,[20] W. Johnson,[8] M. Jones,[51] K.K. Joo,[30] S.Y. Jun,[14] J.E. Jung,[30] T.R. Junk,[20] T. Kamon,[57] D. Kar,[21] P.E. Karchin,[62] Y. Kato[m,44] R. Kephart,[20] W. Ketchum,[65] J. Keung,[48] V. Khotilovich,[57] B. Kilminster,[20] D.H. Kim,[30] H.S. Kim,[30] H.W. Kim,[30] J.E. Kim,[30] M.J. Kim,[22] S.B. Kim,[30] S.H. Kim,[59] Y.K. Kim,[65] N. Kimura,[61] L. Kirsch,[7] S. Klimenko,[21] K. Kondo,[61] D.J. Kong,[30] J. Konigsberg,[21] A. Korytov,[21] A.V. Kotwal,[19] M. Kreps,[29] J. Kroll,[48] D. Krop,[65] N. Krumnack,[5] M. Kruse,[19] V. Krutelyov,[66] T. Kuhr,[29] N.P. Kulkarni,[62] M. Kurata,[59] S. Kwang,[65] A.T. Laasanen,[51] S. Lami,[49] S. Lammel,[20] M. Lancaster,[33] R.L. Lander,[8] K. Lannon[u,42] A. Lath,[55] G. Latino[ff,49] I. Lazzizzera[dd,46] T. LeCompte,[2] E. Lee,[57] H.S. Lee,[65] J.S. Lee,[30] S.W. Lee[w,57] S. Leone,[49] J.D. Lewis,[20] C.-J. Lin,[31] J. Linacre,[45] M. Lindgren,[20] E. Lipeles,[48] A. Lister,[23] D.O. Litvintsev,[20] C. Liu,[50] T. Liu,[20] N.S. Lockyer,[48] A. Loginov,[64] L. Lovas,[17] D. Lucchesi[dd,46] J. Lueck,[29] P. Lujan,[31] P. Lukens,[20] G. Lungu,[53] J. Lys,[31] R. Lysak,[17] D. MacQueen,[36] R. Madrak,[20] K. Maeshima,[20] K. Makhoul,[35] P. Maksimovic,[28] S. Malde,[45] S. Malik,[33] G. Manca[e,32] A. Manousakis-Katsikakis,[3] F. Margaroli,[51] C. Marino,[29] C.P. Marino,[27] A. Martin,[64] V. Martin[k,24] M. Martínez,[4] R. Martínez-Ballarín,[34] P. Mastrandrea,[54] M. Mathis,[28] M.E. Mattson,[62] P. Mazzanti,[6] K.S. McFarland,[52] P. McIntyre,[57] R. McNulty[j,32] A. Mehta,[32] P. Mehtala,[26] A. Menzione,[49] C. Mesropian,[53] T. Miao,[20] D. Mietlicki,[37] N. Miladinovic,[7] R. Miller,[38] C. Mills,[25] M. Milnik,[29] A. Mitra,[1] G. Mitselmakher,[21] H. Miyake,[59] S. Moed,[25] N. Moggi,[6] M.N. Mondragon[n,20] C.S. Moon,[30] R. Moore,[20] M.J. Morello,[49] J. Morlock,[29] P. Movilla Fernandez,[20] J. Mülmenstädt,[31] A. Mukherjee,[20] Th. Muller,[29] P. Murat,[20] M. Mussini[cc,6] J. Nachtman[o,20] Y. Nagai,[59] J. Naganoma,[59] K. Nakamura,[59] Nakano,[43] A. Napier,[60] J. Nett,[63] C. Neu[z,48] M.S. Neubauer,[27] S. Neubauer,[29] J. Nielsen[g,31] L. Nodulman,[2] M. Norman,[10] O. Norniella,[27] E. Nurse,[33] L. Oakes,[45] S.H. Oh,[19] Y.D. Oh,[30] I. Oksuzian,[21] T. Okusawa,[44] R. Orava,[26] K. Osterberg,[26] S. Pagan Griso[dd,46] C. Pagliarone,[58] E. Palencia,[20] V. Papadimitriou,[20] A. Papaikonomou,[29] A.A. Paramanov,[2] B. Parks,[42] S. Pashapour,[36] J. Patrick,[20] G. Pauletta[ii,58] M. Paulini,[14] C. Paus,[35] T. Peiffer,[29] D.E. Pellett,[8] A. Penzo,[58] T.J. Phillips,[19] G. Piacentino,[49] E. Pianori,[48] L. Pinera,[21] K. Pitts,[27] C. Plager,[9] L. Pondrom,[63] K.

Potamianos,[51] O. Poukhov[*,18] F. Prokoshin[y,18] A. Pronko,[20] F. Ptohos[i,20] E. Pueschel,[14] G. Punzi[ee,49] J. Pursley,[63] J. Rademacker[c,45] A. Rahaman,[50] V. Ramakrishnan,[63] N. Ranjan,[51] I. Redondo,[34] P. Renton,[45] M. Renz,[29] M. Rescigno,[54] S. Richter,[29] F. Rimondi[cc,6] L. Ristori,[49] A. Robson,[24] T. Rodrigo,[13] T. Rodriguez,[48] E. Rogers,[27] S. Rolli,[60] R. Roser,[20] M. Rossi,[58] R. Rossin,[66] P. Roy,[36] A. Ruiz,[13] J. Russ,[14] V. Rusu,[20] B. Rutherford,[20] H. Saarikko,[26] A. Safonov,[57] W.K. Sakumoto,[52] L. Santi[ii,58] L. Sartori,[49] K. Sato,[59] A. Savoy-Navarro,[47] P. Schlabach,[20] A. Schmidt,[29] E.E. Schmidt,[20] M.A. Schmidt,[65] M.P. Schmidt[*,64] M. Schmitt,[41] T. Schwarz,[8] L. Scodellaro,[13] A. Scribano[ff,49] F. Scuri,[49] A. Sedov,[51] S. Seidel,[40] Y. Seiya,[44] A. Semenov,[18] L. Sexton-Kennedy,[20] F. Sforza[ee,49] A. Sfyrla,[27] S.Z. Shalhout,[62] T. Shears,[32] P.F. Shepard,[50] M. Shimojima[t,59] S. Shiraishi,[65] M. Shochet,[65] Y. Shon,[63] I. Shreyber,[39] A. Simonenko,[18] P. Sinervo,[36] A. Sisakyan,[18] A.J. Slaughter,[20] J. Slaunwhite,[42] K. Sliwa,[60] J.R. Smith,[8] F.D. Snider,[20] R. Snihur,[36] A. Soha,[20] S. Somalwar,[55] V. Sorin,[4] P. Squillacioti[ff,49] M. Stanitzki,[64] R. St. Denis,[24] B. Stelzer,[36] O. Stelzer-Chilton,[36] D. Stentz,[41] J. Strologas,[40] G.L. Strycker,[37] J.S. Suh,[30] A. Sukhanov,[21] I. Suslov,[18] A. Taffard[f,27] R. Takashima,[43] Y. Takeuchi,[59] R. Tanaka,[43] J. Tang,[65] M. Tecchio,[37] P.K. Teng,[1] J. Thom[h,20] J. Thome,[14] G.A. Thompson,[27] E. Thomson,[48] P. Tipton,[64] P. Ttito-Guzmán,[34] S. Tkaczyk,[20] D. Toback,[57] S. Tokar,[17] K. Tollefson,[38] T. Tomura,[59] D. Tonelli,[20] S. Torre,[22] D. Torretta,[20] P. Totaro[ii,58] S. Tourneur,[47] M. Trovato[gg,49] S.-Y. Tsai,[1] Y. Tu,[48] N. Turini[ff,49] F. Ukegawa,[59] S. Uozumi,[30] N. van Remortel[b,26] A. Varganov,[37] E. Vataga[gg,49] F. Vázquez[n,21] G. Velev,[20] C. Vellidis,[3] M. Vidal,[34] I. Vila,[13] R. Vilar,[13] M. Vogel,[40] I. Volobouev[w,31] G. Volpi[ee,49] P. Wagner,[48] R.G. Wagner,[2] R.L. Wagner,[20] W. Wagner[aa,29] J. Wagner-Kuhr,[29] T. Wakisaka,[44] R. Wallny,[9] S.M. Wang,[1] A. Warburton,[36] D. Waters,[33] M. Weinberger,[57] J. Weinelt,[29] W.C. Wester III,[20] B. Whitehouse,[60] D. Whiteson[f,48] A.B. Wicklund,[2] E. Wicklund,[20] S. Wilbur,[65] G. Williams,[36] H.H. Williams,[67] M.G. Wilson,[56] P. Wilson,[20] B.L. Winer,[42] P. Wittich[h,20] S. Wolbers,[20] C. Wolfe,[65] H. Wolfe,[42] T. Wright,[37] X. Wu,[23] F. Würthwein,[10] A. Yagil,[10] K. Yamamoto,[44] J. Yamaoka,[19] U.K. Yang[r,65] Y.C. Yang,[30] W.M. Yao,[31] G.P. Yeh,[20] K. Yi[o,20] J. Yoh,[20] K. Yorita,[61] T. Yoshida[l,44] G.B. Yu,[19] I. Yu,[30] S.S. Yu,[20] J.C. Yun,[20] A. Zanetti,[58] Y. Zeng,[19] X. Zhang,[27] Y. Zheng[d,9] and S. Zucchelli[cc6]

(CDF Collaboration[†])

[1]Institute of Physics, Academia Sinica, Taipei, Taiwan 11529, Republic of China
[2]Argonne National Laboratory, Argonne, Illinois 60439
[3]University of Athens, 157 71 Athens, Greece
[4]Institut de Fisica d'Altes Energies, Universitat Autonoma de Barcelona, E-08193, Bellaterra (Barcelona), Spain
[5]Baylor University, Waco, Texas 76798
[6]Istituto Nazionale di Fisica Nucleare Bologna, [cc]University of Bologna, I-40127 Bologna, Italy
[7]Brandeis University, Waltham, Massachusetts 02254
[8]University of California, Davis, Davis, California 95616
[9]University of California, Los Angeles, Los Angeles, California 90024
[10]University of California, San Diego, La Jolla, California 92093
[11]University of California, Santa Barbara, Santa Barbara,California 93106
[12]University of California, Irvine, Irvine, California 92697
[13]Instituto de Fisica de Cantabria, CSIC-University of Cantabria, 39005 Santander, Spain
[14]Carnegie Mellon University, Pittsburgh, PA 15213
[15]Enrico Fermi Institute, University of Chicago, Chicago,Illinois 60637
[16]European Organization for Nuclear Research, Geneva 23, Switzerland
[17]Comenius University, 842 48 Bratislava, Slovakia; Institute of Experimental Physics, 040 01 Kosice, Slovakia
[18]Joint Institute for Nuclear Research, RU-141980 Dubna, Russia
[19]Duke University, Durham, North Carolina 27708
[20]Fermi National Accelerator Laboratory, Batavia, Illinois 60510
[21]University of Florida, Gainesville, Florida 32611
[22]Laboratori Nazionali di Frascati, Istituto Nazionale di Fisica Nucleare, I-00044 Frascati, Italy
[23]University of Geneva, CH-1211 Geneva 4, Switzerland
[24]Glasgow University, Glasgow G12 8QQ, United Kingdom
[25]Harvard University, Cambridge, Massachusetts 02138
[26]Division of High Energy Physics, Department of Physics,
University of Helsinki and Helsinki Institute of Physics, FIN-00014, Helsinki, Finland
[27]University of Illinois, Urbana, Illinois 61801
[28]The Johns Hopkins University, Baltimore, Maryland 21218
[29]Institut fur Experimentelle Kernphysik, Karlsruhe Institute of Technology, D-76131 Karlsruhe, Germany
[30]Center for High Energy Physics: Kyungpook National University,
Daegu 702-701, Korea; Seoul National University, Seoul 151-742,
Korea; Sungkyunkwan University, Suwon 440-746,
Korea; Korea Institute of Science and Technology Information,

Daejeon 305-806, Korea; Chonnam National University, Gwangju 500-757,
Korea; Chonbuk National University, Jeonju 561-756, Korea
[31]Ernest Orlando Lawrence Berkeley National Laboratory, Berkeley, California 94720
[32]University of Liverpool, Liverpool L69 7ZE, United Kingdom
[33]University College London, London WC1E 6BT, United Kingdom
[34]Centro de Investigaciones Energeticas Medioambientales y Tecnologicas, E-28040 Madrid, Spain
[35]Massachusetts Institute of Technology, Cambridge, Massachusetts 02139
[36]Institute of Particle Physics: McGill University, Montreal, Quebec,
Canada H3A 2T8; Simon Fraser University, Burnaby, British Columbia,
Canada V5A 1S6; University of Toronto, Toronto, Ontario,
Canada M5S 1A7; and TRIUMF, Vancouver, British Columbia, Canada V6T 2A3
[37]University of Michigan, Ann Arbor, Michigan 48109
[38]Michigan State University, East Lansing, Michigan 48824
[39]Institution for Theoretical and Experimental Physics, ITEP, Moscow 117259, Russia
[40]University of New Mexico, Albuquerque, New Mexico 87131
[41]Northwestern University, Evanston, Illinois 60208
[42]The Ohio State University, Columbus, Ohio 43210
[43]Okayama University, Okayama 700-8530, Japan
[44]Osaka City University, Osaka 588, Japan
[45]University of Oxford, Oxford OX1 3RH, United Kingdom
[46]Istituto Nazionale di Fisica Nucleare, Sezione di Padova-Trento, [dd]University of Padova, I-35131 Padova, Italy
[47]LPNHE, Universite Pierre et Marie Curie/IN2P3-CNRS, UMR7585, Paris, F-75252 France
[48]University of Pennsylvania, Philadelphia, Pennsylvania 19104
[49]Istituto Nazionale di Fisica Nucleare Pisa, [ee]University of Pisa,
[ff]University of Siena and [gg]Scuola Normale Superiore, I-56127 Pisa, Italy
[50]University of Pittsburgh, Pittsburgh, Pennsylvania 15260
[51]Purdue University, West Lafayette, Indiana 47907
[52]University of Rochester, Rochester, New York 14627
[53]The Rockefeller University, New York, New York 10021
[54]Istituto Nazionale di Fisica Nucleare, Sezione di Roma 1,
[hh]Sapienza Universit_a di Roma, I-00185 Roma, Italy
[55]Rutgers University, Piscataway, New Jersey 08855
[56]SLAC National Accelerator Laboratory, Menlo Park, California 94025
[57]Texas A&M University, College Station, Texas 77843
[58]Istituto Nazionale di Fisica Nucleare Trieste/Udine,
I-34100 Trieste, [ii]University of Trieste/Udine, I-33100 Udine, Italy
[59]University of Tsukuba, Tsukuba, Ibaraki 305, Japan
[60]Tufts University, Medford, Massachusetts 02155
[61]Waseda University, Tokyo 169, Japan
[62]Wayne State University, Detroit, Michigan 48201
[63]University of Wisconsin, Madison, Wisconsin 53706
[64]Yale University, New Haven, Connecticut 06520
[65]Enrico Fermi Institute, University of Chicago, Chicago, Illinois 60637
[66]University of California, Santa Barbara, Santa Barbara, California 93106
[67]University of Pennsylvania, Philadelphia,Pennsylvania 19104

[*] Deceased

[†]With visitors from [a]University of Massachusetts Amherst, Amherst, Massachusetts 01003, [b]Universiteit Antwerpen, B-2610 Antwerp, Belgium, [c]University of Bristol, Bristol BS8 1TL, United Kingdom, [d]Chinese Academy of Sciences, Beijing 100864, China, [e]Istituto Nazionale di Fisica Nucleare, Sezione di Cagliari, 09042 Monserrato (Cagliari), Italy, [f] University of California Irvine, Irvine, CA 92697, [g]University of California Santa Cruz, Santa Cruz, CA 95064, [h]Cornell University, Ithaca, NY 14853, [i]University of Cyprus, Nicosia CY-1678, Cyprus, [j]University Col-lege Dublin, Dublin 4, Ireland, [k]University of Edinburgh, Edin-burgh EH9 3JZ, United Kingdom, [l]University of Fukui, Fukui City, Fukui Prefecture, Japan 910-0017 [m]Kinki University, Higashi-Osaka City, Japan 577-8502 [n]Universidad Iberoamericana, Mexico D.F., Mexico, [o]University of Iowa, Iowa City, IA 52242, [p]Kansas State University, Manhattan, KS 66506 [q]Queen Mary, University of London, London, E1 4NS, England, [r]University of Manchester, Manchester M13 9PL, England, [s]Muons, Inc., Batavia, IL 60510, [t]Nagasaki Institute of Applied Science, Nagasaki, Japan,[u]University of Notre Dame, Notre Dame, IN 46556, [v]University de Oviedo, E-33007 Oviedo, Spain, [w]Texas Tech University, Lubbock, TX 79609, [x]IFIC(CSIC-Universitat de Valencia), 56071 Valencia, Spain, [y]Universidad Tecnica Federico Santa Maria, 110v Valparaiso, Chile, [z]University of Virginia, Charlottesville, VA 22906 [aa]Bergische Universit☐at Wuppertal, 42097 Wuppertal, Germany, [bb]Yarmouk University, Irbid 211-

63, Jordan [jj]On leave from J. Ste-fan Institute, Ljubljana, Slovenia,

*March 7, 2010*

**Abstract**

We study the underlying event in proton-antiproton collisions by examining the behavior of charged particles (transverse momentum $p_T > 0.5$ GeV/c, pseudorapidity $|\eta| < 1$) produced in association with large transverse momentum jets (~2.2 $fb^{-1}$) or with Drell-Yan lepton-pairs (~2.7 $fb^{-1}$) in the Z-boson mass region (70 < M(pair) < 110 GeV/$c^2$) as measured by CDF at 1.96 TeV center-of-mass energy. We use the direction of the lepton-pair (in Drell-Yan production) or the leading jet (in high-$p_T$ jet production) in each event to define three regions of $\eta$-$\phi$ space; toward, away, and transverse, where $\phi$ is the azimuthal scattering angle. For Drell-Yan production (excluding the leptons) both the toward and transverse regions are very sensitive to the underlying event. In high-$p_T$ jet production the transverse region is very sensitive to the underlying event and is separated into a MAX and MIN transverse region, which helps separate the hard component (initial and final-state radiation) from the beam-beam remnant and multiple parton interaction components of the scattering. The data are corrected to the particle level to remove detector effects and are then compared with several QCD Monte-Carlo models. The goal of this analysis is to provide data that can be used to test and improve the QCD Monte-Carlo models of the underlying event that are used to simulate hadron-hadron collisions.

# I. INTRODUCTION

In order to find physics beyond the Standard Model at a hadron-hadron collider, it is essential to have Monte-Carlo models that accurately simulate QCD hard-scattering events. To do this one must not only have a good model of the hard-scattering part of the process, but also of the beam-beam remnants (BBR) and the multiple parton interactions (MPI). The underlying event consists of the BBR plus MPI and is an unavoidable background to most collider observables. A good understanding of the underlying event will lead to more precise measurements at the Tevatron and the Large Hadron Collider (LHC). The goal of this analysis is to provide data that can be used to test and improve the QCD Monte Carlo models of the underlying event.

Figure 1.1 illustrates the way the QCD Monte-Carlo models simulate a proton-antiproton collision in which a hard 2-to-2 parton scattering with transverse momentum, $p_T$(hard), has occurred. The resulting event contains particles that originate from the two outgoing partons (plus initial and final-state radiation) and particles that come from the breakup of the proton and antiproton. The beam-beam remnants are what is left over after a parton is knocked out of each of the initial two beam hadrons. They are one of the reasons why hadron-hadron collisions are more complicated than electron-positron annihilations. For the QCD Monte-Carlo models the beam-beam remnants are an important component of the underlying event. In addition to the hard 2-to-2 parton-parton scattering and the beam-beam remnants, sometimes there are additional semi-hard 2-to-2 parton-parton scatterings (MPI) that contribute particles to the underlying event. However, on an event-by-event basis these two components cannot be uniquely separated from particles that come from the initial and final-state radiation. Hence, a study of the underlying event inevitably involves a study of the BBR plus MPI plus initial and final-state radiation.

As shown in Fig. 1.2, Drell-Yan lepton-pair production provides an excellent place to study the underlying event. Here one studies the outgoing charged particles (excluding the lepton pair) as a function of the lepton-pair invariant mass and as a function of the lepton-pair transverse momentum. Unlike high-$p_T$ jet production, for lepton-pair production there is no final-state gluon radiation.

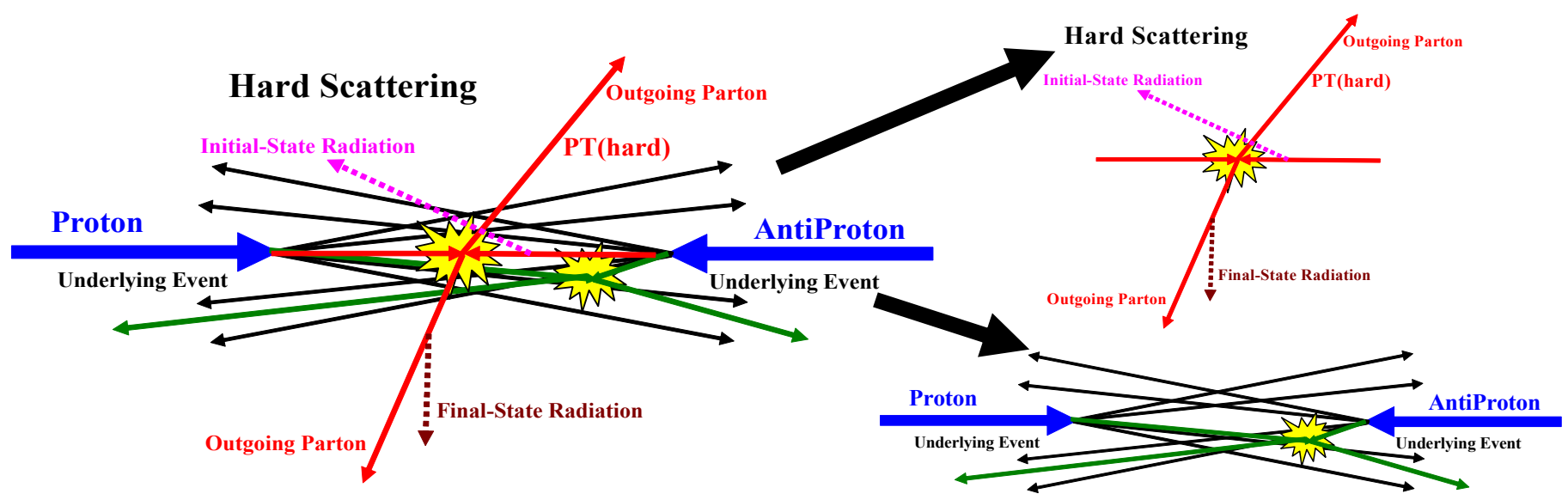

**Fig. 1.1**. Illustration of the way QCD Monte-Carlo models simulate a proton-antiproton collision in which a hard 2-to-2 parton scattering with transverse momentum, $p_T$(hard), has occurred. The hard-scattering component of the event consists of particles that result from the hadronization of the two outgoing partons (*i.e.* the initial two jets) plus the particles that arise from initial and final state radiation (*i.e.* multi-jets). The underlying event consists of particles that arise from the beam-beam remnants and from multiple parton interactions.

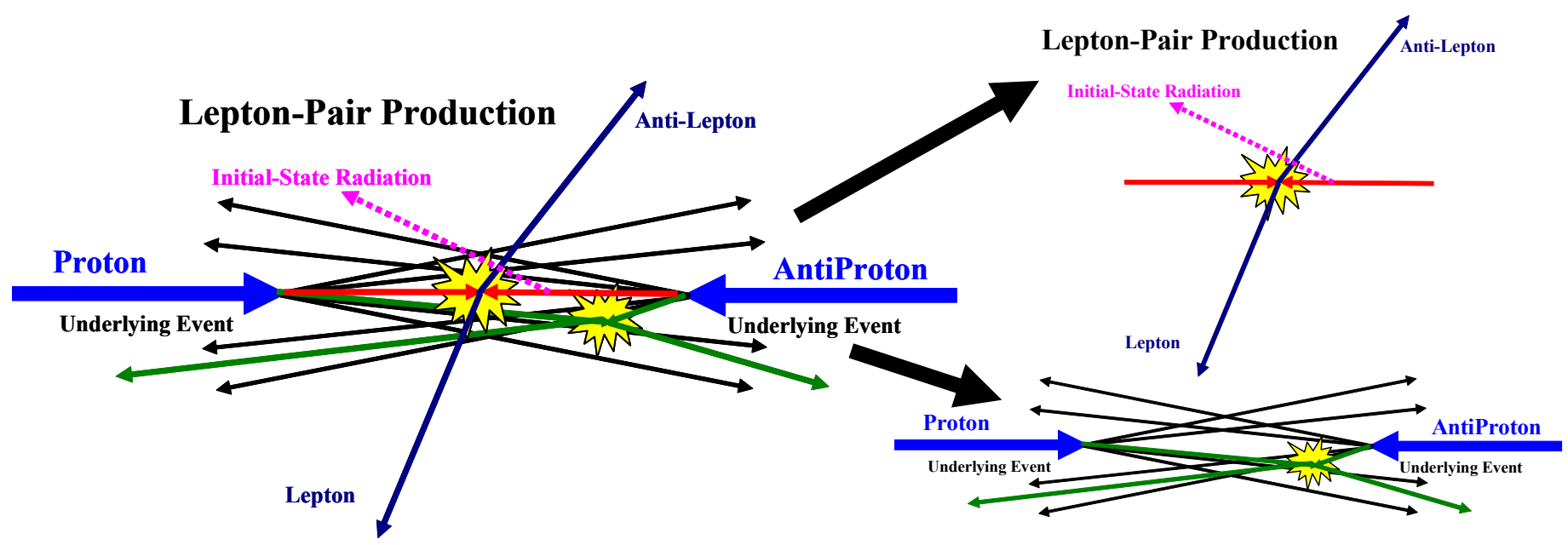

**Fig. 1.2**. Illustration of the way QCD Monte-Carlo models simulate Drell-Yan lepton-pair production. The hard-scattering component of the event consists of the two outgoing leptons plus particles that result from initial-state radiation. The underlying event consists of particles that arise from the beam-beam remnants and from multiple parton interactions.

Hard-scattering collider jet events have a distinct topology. A typical hard-scattering event consists of a collection (or burst) of hadrons traveling roughly in the direction of the initial two beam particles and two collections of hadrons (jets) with large transverse momentum. The two large transverse momentum jets are roughly back-to-back in azimuthal angle, $\phi$. One can use the topological structure of hadron-hadron collisions to study the underlying event. We use the direction of the leading (highest $p_T$) jet in each event to define four regions of $\eta$-$\phi$ space, where $\eta$ is the pseudorapidity. The pseudorapidity $\eta = -\log(\tan(\theta_{cm}/2))$, where $\theta_{cm}$ is the center-of-mass polar scattering angle and $\phi$ is the azimuthal angle of outgoing charged particles. As illustrated in Fig. 1.3, the direction of the leading jet, jet#1, in high-$p_T$ jet production or the direction of the lepton-pair in Drell-Yan production is used to define correlations in the azimuthal angle, $\Delta\phi$. The angle $\Delta\phi = \phi - \phi_{jet\#1}$ ($\Delta\phi = \phi - \phi_{pair}$) is the relative azimuthal angle between a charged particle and the direction of jet#1 (lepton-pair). The toward region is defined by $|\Delta\phi| < 60^o$ and $|\eta| < 1$, while the away region is $|\Delta\phi| > 120^o$ and $|\eta| < 1$. The two transverse regions $60^o < -\Delta\phi < 120^o$, $|\eta| < 1$ and $60^o < \Delta\phi < 120^o$, $|\eta| < 1$ are referred to as transverse 1 and transverse 2. The overall transverse region corresponds to combining the transverse-1 and transverse-2 regions. In high-$p_T$ jet production, the toward and away regions receive large contributions from the outgoing high-$p_T$ jets, while the transverse region is perpendicular to the plane of the hard 2-to-2 scattering and is therefore very sensitive to the underlying event. For Drell-Yan production both the toward and the transverse region are very sensitive to the underlying event, while the away region receives large contributions from the away-side jet from the subprocesses: $q+\bar{q}\rightarrow l^+l^- + g$ , $q+g\rightarrow l^+l^- + q$ , $\bar{q}+g\rightarrow l^+l^- + \bar{q}$ .

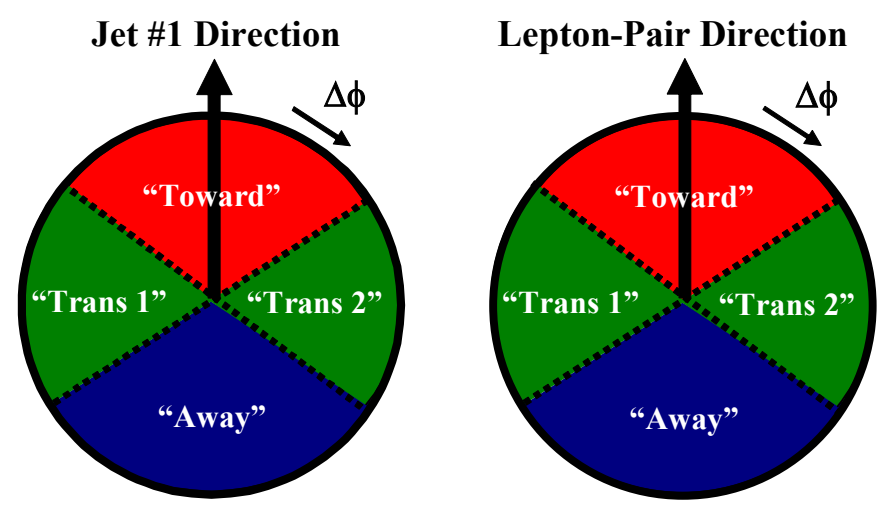

**Fig. 1.3**. Illustration of correlations in azimuthal angle $\Delta\phi$ relative to (*left*) the direction of the leading jet (highest $p_T$ jet) in the event, jet#1, in high-$p_T$ jet production or (*right*) the direction of the lepton-pair in Drell-Yan production. The angle $\Delta\phi = \phi - \phi$jet#1 ($\Delta\phi = \phi - \phi_{pair}$) is the relative azimuthal angle between charged particles and the direction of jet#1 (lepton-pair). The toward region is defined by $|\Delta\phi| < 60^o$ and $|\eta| < 1$, while the away region is $|\Delta\phi| > 120^o$ and $|\eta| < 1$. The two transverse regions $60^o < -\Delta\phi < 120^o$, $|\eta| < 1$ and $60^o < \Delta\phi < 120^o$, $|\eta| < 1$ are referred to as transverse 1 and transverse 2. Each of the two transverse regions have an area in $\eta$-$\phi$ space of $\Delta\eta\Delta\phi = 4\pi/6$. The overall transverse region corresponds to combining the transverse-1 and transverse-2 regions. The transMAX (transMIN) refer to the transverse region (transverse-1 or transverse-2) containing the largest (smallest) number of charged particles or to the region containing the largest (smallest) scalar $p_T$ sum of charged particles

**Table 1.1**. Observables examined in this analysis as they are defined at the particle level and the detector level. Charged tracks are considered good if they pass the track selection criterion given in Section III(5). The mean charged-particle $<p_T>$ is constructed on an event-by-event basis and then averaged over the events. For the average $p_T$ and the $PT_{max}$, we require that there is at least one charged particle present. Particles are considered stable if $c\tau > 10$ mm ($K_s$, $\Lambda$, $\Sigma$, $\Xi$, and $\Omega$ are kept stable) .

| Observable | Particle Level | Detector level |
|---|---|---|
| $dN/d\eta d\phi$ | Number of stable charged particles per unit $\eta$-$\phi$ ($p_T > 0.5$ GeV/c, $\|\eta\| < 1$) | Number of good tracks per unit $\eta$-$\phi$ ($p_T > 0.5$ GeV/c, $\|\eta\| < 1$) |
| $dPT/d\eta d\phi$ | Scalar $p_T$ sum of stable charged particles per unit $\eta$-$\phi$ ($p_T > 0.5$ GeV/c, $\|\eta\| < 1$) | Scalar $p_T$ sum of good tracks per unit $\eta$-$\phi$ ($p_T > 0.5$ GeV/c, $\|\eta\| < 1$) |
| $<p_T>$ | Average $p_T$ of stable charged particles ($p_T > 0.5$ GeV/c, $\|\eta\| < 1$) Require at least 1 charged particle | Average $p_T$ of good tracks ($p_T > 0.5$ GeV/c, $\|\eta\| < 1$) Require at least 1 good track |
| $PT_{max}$ | Maximum $p_T$ stable charged particle ($p_T > 0.5$ GeV/c, $\|\eta\| < 1$) Require at least 1 charged particle | Maximum $p_T$ good charged tracks ($p_T > 0.5$ GeV/c, $\|\eta\| < 1$) Require at least 1 good track |
| Jet | MidPoint algorithm R = 0.7 $f_{merge}$ = 0.75 applied to stable particles | MidPoint algorithm R = 0.7 $f_{merge}$ = 0.75 applied to calorimeter cells |

We study charged particles in the range $p_T > 0.5$ GeV/c and $|\eta| < 1$ in the toward, away and transverse regions. For high-$p_T$ jet production, we require that the leading jet in the event be in the region $|\eta(\text{jet\#1})| < 2$ (referred to as leading-jet events). The jets are constructed using the MidPoint algorithm ($R = 0.7$, $f_{merge} = 0.75$), where R is the jet radius and $f_{merge}$ is the jet splitting and merging fraction [1]. For Drell-Yan production we require that the invariant mass of the lepton-pair be in the mass region of the Z-boson, $70 < M(\text{pair}) < 110$ GeV/c$^2$, with $|\eta(\text{pair})| < 6$ (referred to as Drell-Yan events). For both leading-jet and Drell-Yan events we define MAX and MIN transverse regions (transMAX and transMIN) [2], where MAX (MIN) refers to the transverse region (transverse-1 or transverse-2) containing the largest (smallest) number of charged particles or to the region containing the largest (smallest) scalar $p_T$ sum of charged particles. For events with large initial or final-state radiation the transMAX region will usually contain the third jet in high-$p_T$ jet production or the second jet in Drell-Yan production while both the transMAX and transMIN regions receive contributions from the beam-beam remnants. Thus, the transMIN region is very sensitive to the beam-beam remnants, while the event-by-event difference between transMAX and transMIN is very sensitive to initial and final-state radiation (transDIF = transMAX – transMIN).

Table 1.1 shows the observables that are considered in this analysis as they are defined at the particle level and detector level. The detector level corresponds to the tracks passing good-track criteria and the particle level corresponds to true charged particles in the event. The particle level can be compared directly with the QCD Monte-Carlo models at the generator level. Since we will be studying regions in $\eta$-$\phi$ space with different areas, we construct densities by dividing by the area. For example, the number density $dN/d\eta d\phi$, corresponds the number of charged particles per unit $\eta$-$\phi$ and the $PT_{sum}$ density $dPT/d\eta d\phi$, corresponds the charged-particle scalar-$p_T$ sum per unit $\eta$-$\phi$.

A discussion of the QCD Monte-Carlo Model is presented in Section II. In Section III we discuss the data selection, track cuts, and the method we use to correct the data to the particle level. Section IV contains the results for leading-jet and Drell-Yan events and comparisons with the QCD Monte-Carlo models. Section V is reserved for the summary and conclusions.

# II. QCD Monte-Carlo Models

QCD Monte-Carlo generators such as PYTHIA [3] have parameters which may be adjusted to control the behavior of their event modeling. A specified set of these parameters that has been adjusted to better fit some aspects of the data is referred to as a tune. PYTHIA Tune A was determined by fitting the CDF Run 1 underlying event data [4]. Later it was noticed that Tune A does not fit the CDF Run 1 Z-boson $p_T$ distribution very well [5]. PYTHIA Tune AW fits the Z-boson $p_T$ distribution as well as the underlying event at the Tevatron [6]. For leading-jet production Tune A and Tune AW are nearly identical. Table 2.1 shows the parameters for several tunes for PYTHIA version 6.2. PYTHIA Tune DW is very similar to Tune AW except the setting of one PYTHIA parameter PARP(67) = 2.5, which is the preferred value determined by the DØ Collaboration in fitting their dijet $\Delta\phi$ distribution [7]. PARP(67) sets the high-$p_T$ scale for initial-state radiation in PYTHIA. It determines the maximal parton virtuality allowed in time-like showers. Tune DW and Tune DWT are identical at 1.96 TeV (the reference point), but Tune DW and DWT extrapolate differently to the LHC. Tune DWT uses the PYTHIA default value for energy dependence of the MPI cut-off (PARP(90) = 0.16), which is the value used in the ATLAS PYTHIA tune [8]. Tune DWT produces more activity in the underlying event at the LHC than does Tune DW, but predicts less activity than Tune DW in the underlying event at energies below 1.96 TeV. Tune DW uses the Tune A value of PARP(90) = 0.25, which was determined by comparing the Run 1 data at 1.8 TeV with the CDF underlying event measurements at 630 GeV [9]. The amount of MPI and hence the tunings depend on the choice of the parton distribution functions. All these tunes use the CTEQ5L [10] parton distribution functions.

The first 9 parameters in Table 2.1 tune the MPI. PARP(62), PARP(64), and PARP(67) tune the initial-state radiation and the last three parameters set the intrinsic transverse momentum of the partons within the incoming proton and antiproton.

**Table 2.1.** Parameters for several PYTHIA 6.2 tunes. Tune A is the CDF Run 1 underlying-event tune. Tune AW and DW are CDF Run 2 tunes which fit the existing Run 2 underlying event data and fit the Run 1 Z-boson $p_T$ distribution. The ATLAS Tune is the tune used in the ATLAS TDR [8]. Tune DWT uses the ATLAS energy dependence for the MPI, PARP(90). The first 9 parameters tune the multiple parton interactions. PARP(62), PARP(64), and PARP(67) tune the initial-state radiation and the last three parameters set the intrinsic $k_T$ of the partons within the incoming proton and antiproton.

| Parameter | Description | Tune A | Tune AW | Tune DW | Tune DWT | ATLAS |
|---|---|---|---|---|---|---|
| **PDF** | Parton Distribution Functions | CTEQ5L | CTEQ5L | CTEQ5L | CTEQ5L | CTEQ5L |
| **MSTP(81)** | MPI On | 1 | 1 | 1 | 1 | 1 |
| **MSTP(82)** | Double Gaussian Matter Distribution | 4 | 4 | 4 | 4 | 4 |
| **PARP(82)** | MPI Cut-Off | 2.0 | 2.0 | 1.9 | 1.9409 | 1.8 |
| **PARP(83)** | Fraction of matter within core | 0.5 | 0.5 | 0.5 | 0.5 | 0.5 |
| **PARP(84)** | Core Radius | 0.4 | 0.4 | 0.4 | 0.4 | 0.5 |
| **PARP(85)** | Color Connections | 0.9 | 0.9 | 1.0 | 1.0 | 0.33 |
| **PARP(86)** | Color Connections | 0.95 | 0.95 | 1.0 | 1.0 | 0.66 |
| **PARP(89)** | Reference Energy | 1800 | 1800 | 1800 | 1960 | 1000 |
| **PARP(90)** | MPI Energy Dependence | 0.25 | 0.25 | 0.25 | 0.16 | 0.16 |
| **PARP(62)** | Initial-state radiation Cut-Off | 1.0 | 1.25 | 1.25 | 1.25 | 1.0 |
| **PARP(64)** | Soft Initial-State Radiation Scale | 1.0 | 0.2 | 0.2 | 0.2 | 1.0 |
| **PARP(67)** | Hard Initial-State Radiation Scale | 4.0 | 4.0 | 2.5 | 2.5 | 1.0 |
| **MSTP(91)** | Gaussian Intrinsic $k_T$ | 1 | 1 | 1 | 1 | 1 |
| **PARP(91)** | Intrinsic Gaussian Width, $\sigma$ | 1.0 | 2.1 | 2.1 | 2.1 | 1.0 |
| **PARP(93)** | Intrinsic $k_T$ Upper Cut-Off | 5.0 | 15.0 | 15.0 | 15.0 | 5.0 |

**Table 2.2.** The computed value of the multiple parton scattering cross section for the various PYTHIA 6.2 tunes.

| Tune | σ(MPI) at 1.96 TeV | σ(MPI) at 14 TeV |
|---|---|---|
| **A, AW** | 309.7 mb | 484.0 mb |
| **DW** | 351.7 mb | 549.2 mb |
| **DWT** | 351.7 mb | 829.1 mb |
| **ATLAS** | 324.5 mb | 768.0 mb |

Table 2.2 shows the computed value of the multiple parton scattering cross section for the various tunes. The multiple parton scattering cross section (divided by the total inelastic cross section at the center-of-mass energies of 1.96 and 14 Tev, respectively) determines the average number of multiple parton collisions per event. The MPI cross section is the same for proton-proton and proton-antiproton collisions.

HERWIG [11] is a QCD Monte-Carlo generator similar to PYTHIA except HERWIG employs a cluster fragmentation model while PYTHIA uses a string fragmentation approach. In addition, gluon radiation is modeled differently by the two generators. Also, HERWIG does not include MPI in the underlying event. In HERWIG the underlying event arises solely from the BBR. JIMMY [12] is a multiple parton interaction model which can be added to HERWIG to improve agreement with the underlying event observables. To compare with the Drell-Yan data we have constructed a HERWIG tune (with JIMMY MPI) with JMUEO = 1, PTJIM = 3.6 GeV/c,

JMRAD(73) = 1.8, and JMRAD(91) = 1.8. These parameters govern the MPI activity produced by JIMMY. This tune of JIMMY was arrived at by fitting the data from this analysis on the charged scalar particle PTsum density in the toward region for Drell-Yan production.

In this paper the theory predictions are presented as smooth curves. These curves come from fits to QCD Monte-Carlo output with limited statistical accuracy. The theory curves presented here reproduce the QCD Monte-Carlo results (with infinite statistical accuracy) within about 2%.

# III. ANALYSIS STRATEGY

## (1) Data Sample and Event Selection

The CDF Run II detector, in operation since 2001, is an azimuthally and forward-backward symmetric solenoidal particle detector [13]. It combines precision charged particle tracking with fast projective calorimetry and fine grained muon detection. Tracking systems are designed to detect charged particles and measure their momenta and displacements from the point of collision, termed the primary interaction vertex. The tracking system consists of a silicon microstrip system and an open-cell wire drift chamber, termed the Central Outer Tracker (COT) that surrounds the silicon. Segmented electromagnetic and hadronic sampling calorimeters surround the tracking system and measure the energy of interacting particles. Particles make showers which deposit energy and are sampled via their ionization. The muon system resides beyond the calorimeters. Muons are minimally ionizing particles and, hence, only deposit small amounts of ionization energy in the material. They are the only particles likely to penetrate both the tracking and five pion absorption lengths of calorimeter steel, and leave tracks in the muon detection system. At CDF the positive z-axis is defined to lie along the incident proton beam direction. The leading-jet data and lepton-pair data corresponds to an integrated luminosity of about 2.2 $fb^{-1}$ and 2.7 $fb^{-1}$, respectively. For both data sets we require one and only one primary vertex within the fiducial region $|Z_{vertex}| \leq 60$ cm centered around the nominal CDF z =0.

## (2) Jet Selection

Jets are selected using the MidPoint cone based algorithm with a cone size of 0.7 and $f_{merge} = 0.75$ [1]. For the leading-jet events we require that the highest $p_T$ jet in the calorimeter lie in the range $|\eta| < 2$ or the event is rejected.

## (3) Lepton Selection

Dielectron events are triggered online by either one central ($|\eta| < 1.1$) electron candidate with $E_T > 18$ GeV and a track with $p_T > 18$ GeV/c associated to it, or by two electromagnetic clusters with $E_T > 18$ GeV and $|\eta| < 3.2$ where no track association is required. At offline level we consider only electrons with $E_T > 20$ GeV and $|\eta| < 1$ that also have a track matched to the calorimeter cluster. The electrons also have to pass certain quality criteria to verify that they are consistent with the electromagnetic shower characteristics as expected for electrons [14].

Dimuon events are triggered on at least one muon candidate that has a signal in one of the muon chambers with $|\eta| < 1$and $p_T > 18$ GeV/c. The second muon candidate is not required to have a signal in the muon chambers but it must have hits in the COT. At offline level we consider only muon candidates with $p_T > 20$ GeV and $|\eta| < 1$. All muon candidates are required to have calorimeter energy deposits consistent with those expected from a minimum ionizing particle. In addition, we employ a time-of-flight filter to remove cosmic ray muons.

All leptons are required to be isolated from other charged tracks in the event by a distance of $R = \sqrt{(\Delta\eta)^2 + (\Delta\phi)^2} < 0.4$.

## (4) Lepton-Pair Selection

The lepton pairs are formed by oppositely charged leptons, with the requirement that the z positions of the two leptons satisfy $|\Delta z| < 4$ cm, to ensure that both leptons came from the same primary collision. For the Drell-Yan data we require that both leptons have $p_T > 20$ GeV/c and $|\eta| < 1$ and that the invariant mass of the lepton-pair be in the range $70 < M(\text{pair}) < 110$ GeV/c$^2$, with $|\eta(\text{pair})| < 6$. We chose this lepton-pair mass region because studies have shown that the lepton-pair backgrounds (mostly from events with QCD jets or events with a W-boson and jets) are negligible in the region of the Z-boson [15].

## (5) Track Selection

We consider charged tracks that have been measured by the central outer tracker (COT). The COT [16] is a cylindrical open-cell drift chamber with 96 sense wire layers grouped into eight alternating superlayers of stereo and axial wires. Its active volume covers $40 < r < 137$ cm, where r is the radial coordinate in the plane transverse to the z axis, and $|z| < 155$ cm, thus providing fiducial coverage in $|\eta| \leq 1.1$ to tracks originating within $|z| \leq 60$ cm. We include tracks in the region $0.5 < p_T < 150$ GeV/c and $|\eta| < 1$ where COT efficiency is high. At very high $p_T$ the track resolution deteriorates. The upper limit of 150 GeV/c is chosen to prevent mis-measured tracks with very high $p_T$ from distorting the average charged-particle density and the average charged-particle PTsum density. Tracks are required to hit at least two axial segments with more than 10 total hits and at least two stereo segments with more than 10 total hits in the COT. In addition, the tracks are required to point back to the primary vertex. We consider two track selections; loose and tight. The loose track selection requires $|d_0| < 1.0$ cm and $|z - Z_{vtx}| < 3$ cm, where $d_0$ is the beam corrected transverse impact parameter and $z - Z_{vtx}$ is the distance on the z-axis (beam axis) between the track and the primary vertex. The tight track selection requires that $|d_0| < 0.5$ cm and $|z - Z_{vtx}| < 2$ cm. The loose criterion is similar to the Run 1 underlying event analysis [4].

## (6) Correcting to the Particle Level and Systematic Uncertainties

The raw data at the detector level are corrected to the level of final state stable particles and are then compared with the QCD Monte-Carlo models at the generator level. The particle level corresponds to the event without detector effects. The detector level corresponds to the tracks passing the good track criterion. We rely on the QCD Monte-Carlo models and the CDF detector simulation CDFSIM (parametrized response of the CDF II detector [17, 18]) to correct the measured tracks back to the stable charged particle level. Particles are considered stable if $c\tau > 10$ mm. The generator level charged particles have $p_T > 0.5$ GeV/c, $|\eta| < 1$, and are kept stable if $c\tau > 10$ mm. Hence, to compare the corrected data with QCD Monte-Carlo model predictions one must keep the $K_{short}$ meson stable as well as the following baryons: $\Lambda$, $\Sigma$, $\Xi$, and $\Omega$.

The QCD Monte-Carlo models are used to calculate the observables in Table 1.1 at the particle level in bins of particle jet#1 $p_T$ (GEN) and at the detector level in bins of calorimeter jet#1 $p_T$ (CDFSIM). GEN refers to the Monte-Carlo model at the generator level and CDFSIM are the GEN particles after detector simulation. The detector-level data in bins of calorimeter jet#1 $p_T$ are corrected by multiplying by the QCD Monte-Carlo correction factor, $F_{cor}$ = GEN/CDFSIM. This is done bin-by-bin for every observable. We refer to the ratio $F_{res}$ = CDFSIM/GEN as the response factor for that observable with the correction factor being the reciprocal. Smooth curves are drawn through the QCD Monte-Carlo predictions at both the generator level (GEN) and the detector level (CDFSIM) to aid in comparing the theory with the data and also to construct the correction factors. This one step correction method simultaneously corrects for mis-measurement of the leading jet transverse momentum (jet energy scale) and for missed and/or fake tracks.

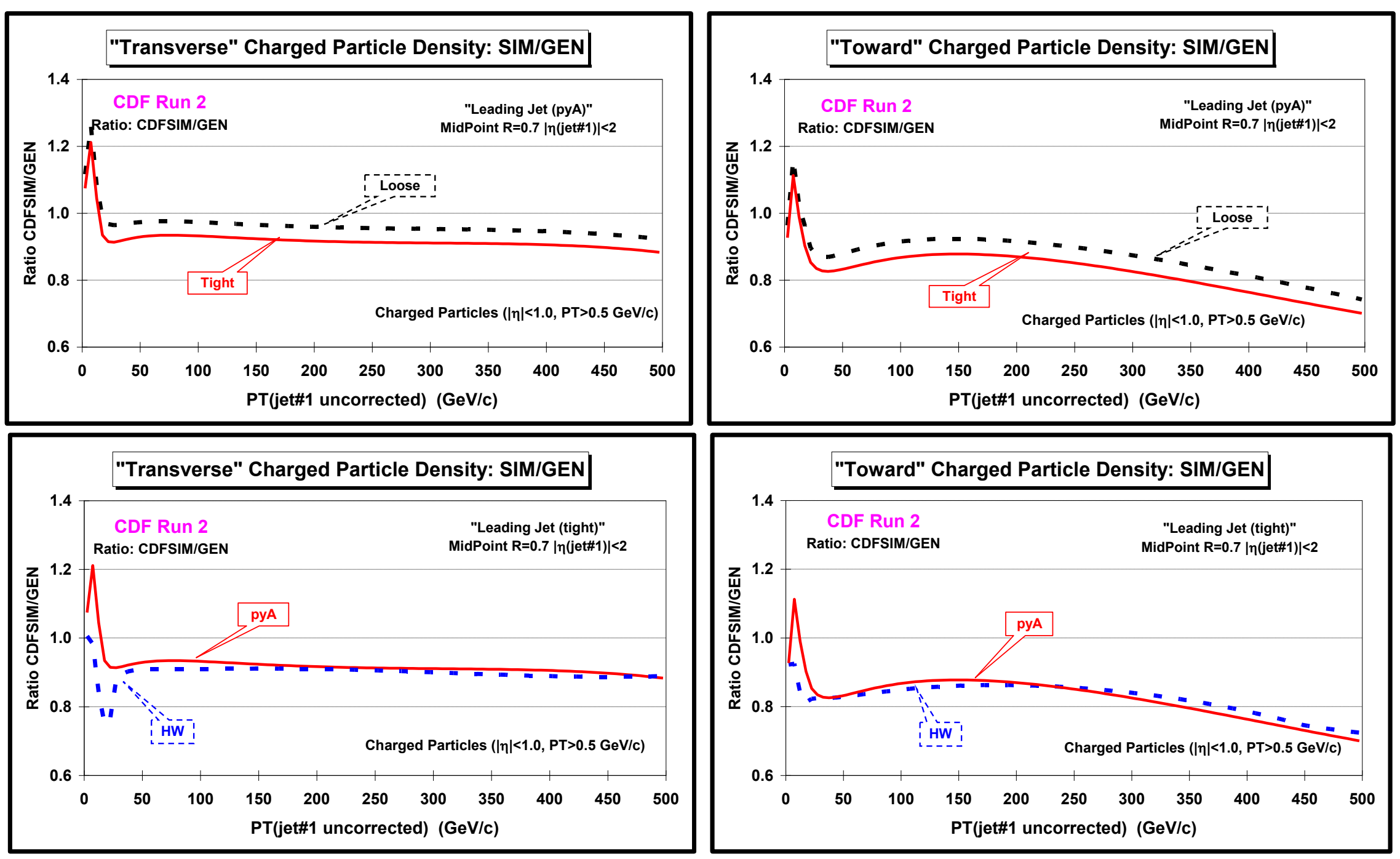

**Fig. 3.1**. The response factors, $F_{res}$ = CDFSIM/GEN, for the charged-particle density, $dN/d\eta d\phi$, in the toward and transverse regions for leading-jet events. The top two plots show the response factors for PYTHIA Tune A (pyA) with tight and loose track cuts for the transverse (*top left*) and toward (*top right*) regions. The bottom two plots compare PYTHIA Tune A (pyA) and HERWIG without MPI (HW) for tight track cuts for the transverse (*bottom left*) and toward (*bottom right*) regions. The correction factor is the reciprocal of the response factor ($F_{cor} = 1/F_{res}$).

The correction factors are different for every observable and they are different for the tight and loose track selection criterion. The tight track criterion results in fewer tracks than the loose criterion and hence the Monte-Carlo corrections factors are different. If the Monte-Carlo described the data perfectly and if CDFSIM were exact, then the corrected observable would be identical regardless of the track selection criterion. Using PYTHIA Tune A for the leading-jet events and PYTHIA Tune AW for the Drell-Yan events, we find that the loose and tight track selections do result in nearly the same particle level result for all the observables presented in this analysis. The differences are used as a source of systematic error and are added in quadrature to the statistical errors.

Figure 3.1 shows the response factors, $F_{res}$, for the charged-particle density in the toward and transverse regions for leading-jet events. The correction factors ($1/F_{res}$) are typically small (they differ from one by less that 10%) except in regions where the charged-particle density becomes large, which occurs in the toward and away regions for leading-jet production. The efficiency of detecting charged tracks decreases when the density of tracks becomes large. For this reason we restrict ourselves to the range $p_T$(jet#1) < 200 GeV/c for the toward and away regions, but allow the leading jet transverse momentum to extend to 400 GeV/c in the transverse region. For the leading-jet events we have also used HERWIG (without MPI) as well as PYTHIA Tune A to correct the data to the particle level. We use the differences in the corrected data as an additional source of systematic error (added in quadrature). For low $p_T$(jet#1) the correction factors become large due to the uncertainty in the jet energy scale at low energy. Also, the corrections from HERWIG and PYTHIA Tune A differ significantly in this region. This results in very large systematic errors on the data at low leading-jet transverse momentum.

Another important effect and resultant systematic error arises from the uncertainty in the jet energy scale for $p_T$ of the leading jet. The CDF detector simulation does not reproduce perfectly the response of the calorimeters. The overall systematic uncertainty in the CDF jet energy scale (JES) is a function of the jet $p_T$ [21]. The uncertainty is about 3% at high $p_T$ and increases to around 8% at low $p_T$. After correcting the data to the particle level we shift $P_T$(jet#1) up and down by this additional uncertainty with the bin-by-bin differences in the observables in Table 1.1 used as another systematic error. The JES systematic errors are large in the toward and away region where the observables are varying rapidly with $P_T$(jet#1).

We investigated the dependence of the corrected data to our upper limit of $PT_{max}$(cut) = 150 GeV/c which was applied to all tracks. The sensitivity of the results to this choice of upper limit was checked by changing the upper limit to $PT_{max}(cut) = 1.5 \times ET_{max}(tower)$. Here one looks, on an event-by-event basis, at all the towers in the region $|\eta| < 1$ and sets the maximum $p_T$ track cut to be equal to 1.5 times the $E_T$ of the tower with the largest transverse energy. High $p_T$ mis-measured tracks do not deposit energy in the calorimeter. The two maximum $p_T$ track cut methods produce slightly different correction factors, however, after correcting to the particle level the results are nearly identical. For the leading-jet analysis the differences were used as an additional systematic error.

Although we require one and only one high quality vertex, the observables in Table 1.1 can still be affected by pile-up (more than one proton-antiproton collision in the event). Tracks are required to point back to the primary vertex, but the track observables are affected by pile-up when two vertices overlap. Vertices within about 3 cm of each other merge together as one. In the leading-jet analysis we examined the effects of pile-up by plotting the transverse charged-particle density and the charged-particle PTsum density versus the instantaneous luminosity (with one and only one vertex). As the instantaneous luminosity increases so does the amount of pile-up. We found that these observables did increase slightly with increasing luminosity (roughly linearly). The leading-jet observables in the transverse region are corrected for pile-up by extrapolating to the low luminosity limit. To correct the data, we define a low region, $L_{inst} < 25 \times 10^{30}$ cm$^{-2}$s$^{-1}$ (low), and a high region $L_{inst} > 25 \times 10^{30}$ cm$^{-2}$s$^{-1}$ (high), where $L_{inst}$ is the instantaneous luminosity. On a bin-by-bin basis, the ratio high/low and all/low was constructed, where all = high + low. The ratio high/low was found to be small (usually less than 1%) and could simply have been absorbed into the overall systematic errors. However, in the leading-jet analysis we corrected the data for pile-up by drawing a smooth curve through the ratio all/low and then dividing the data by this ratio. The size of the pile-up correction was then taken as the systematic error in making the correction and added in quadrature with the other systematic errors. For the Drell-Yan analysis, the pile-up corrections were less than 1% and were simply absorbed into the overall systematic errors.

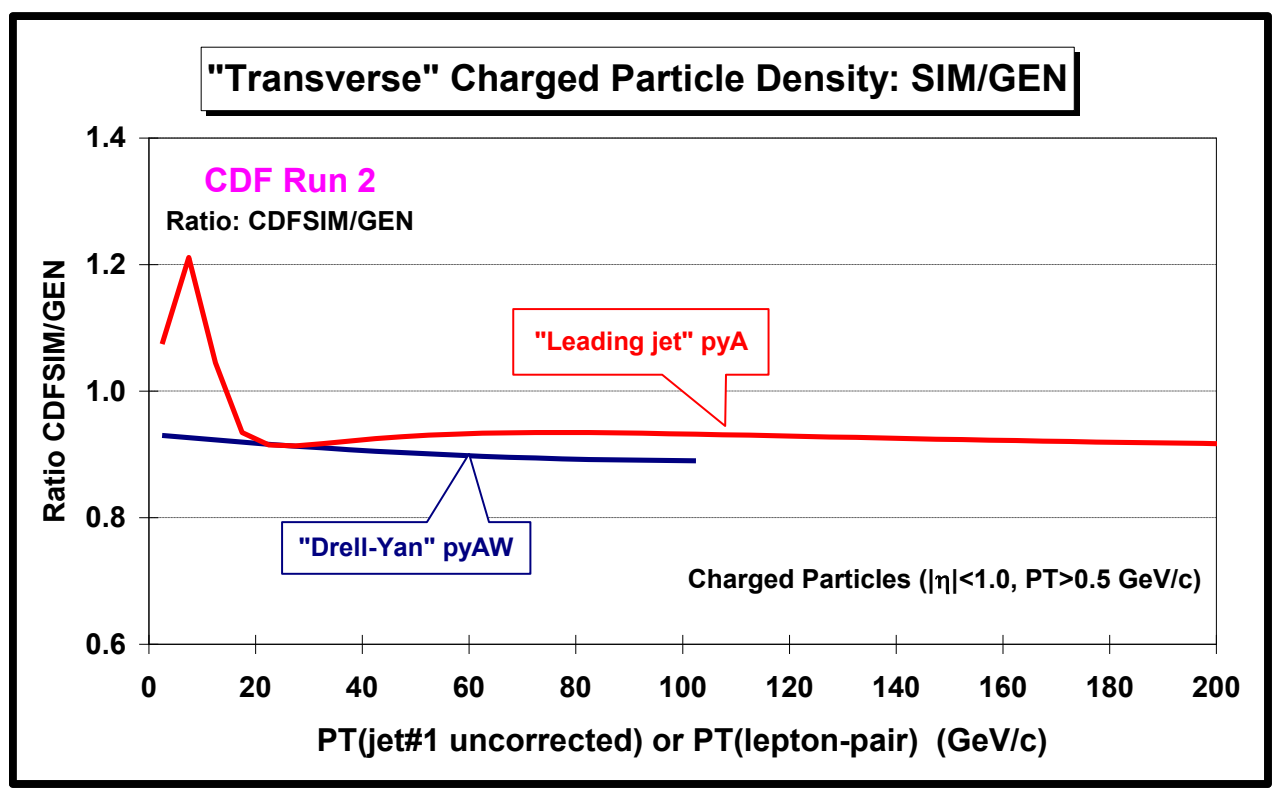

**Fig. 3.2**. The response factors, $F_{res}$ = CDFSIM/GEN, for the charged-particle density, dN/dηdϕ, in the transverse region for leading-jet events and for Drell-Yan events. The plots show the response factors for PYTHIA Tune A (pyA) with tight track cuts (leading-jet) and for PYTHIA Tune AW (pyAW) with tight track cuts (Drell-Yan). The correction factor is the reciprocal of the response factor ($F_{cor} = 1/F_{res}$).

Figure 3.2 shows the response factors, $F_{res}$ = CDFSIM/GEN, for the charged-particle density, dN/dηdϕ, in the transverse region for leading-jet events and for Drell-Yan events. The response factors are similar, but not the same. In the Drell-Yan analysis we required the leptons to be isolated from other particles in the event. This biases one against a very active underlying event which is compensated for by the correction factor.

# IV. RESULTS

## (1) Leading-Jet and Drell-Yan Topologies

Figure 4.1 shows the data on the density of charged particles and the scalar PTsum density, respectively, for the toward, away, and transverse regions for leading-jet and Drell-Yan events. For leading-jet events the densities are plotted as a function of the leading-jet $p_T$ and for Drell-Yan events there are plotted versus the $p_T$ of the lepton pair. The data are corrected to the particle level and are compared with PYTHIA Tune A (leading-jet) and Tune AW (Drell-Yan) at the particle level. For leading-jet events at high $p_T$(jet#1) the densities in the toward and away regions are much larger than in the transverse region because of the toward-side and away-side jets. At small $p_T$(jet#1) the toward, away, and transverse densities become equal and go to zero as $p_T$(jet#1) goes to zero. If the leading jet has no transverse momentum then there can be no particles anywhere. In addition, there are a lot of low transverse momentum jets and for $p_T$(jet#1) < 30 GeV/c the leading jet is not always the jet resulting from the hard 2-to-2 scattering. This produces a bump in the transverse density in the range where the toward, away, and transverse densities become similar in size. For Drell-Yan events the toward and transverse densities are both small and almost equal. The away density is large due to the away-side jet. The toward, away, and transverse densities become equal as $p_T$ of the lepton pair goes to zero, but unlike the leading-jet case the densities do not vanish at $p_T$(lepton-pair) = 0. For Drell-Yan events with $p_T$(lepton-pair) = 0 the hard scale is set by the lepton-pair mass which is in the region of the Z-boson, whereas in leading-jet events the hard scale goes to zero as transverse momentum of the leading jet goes to zero.

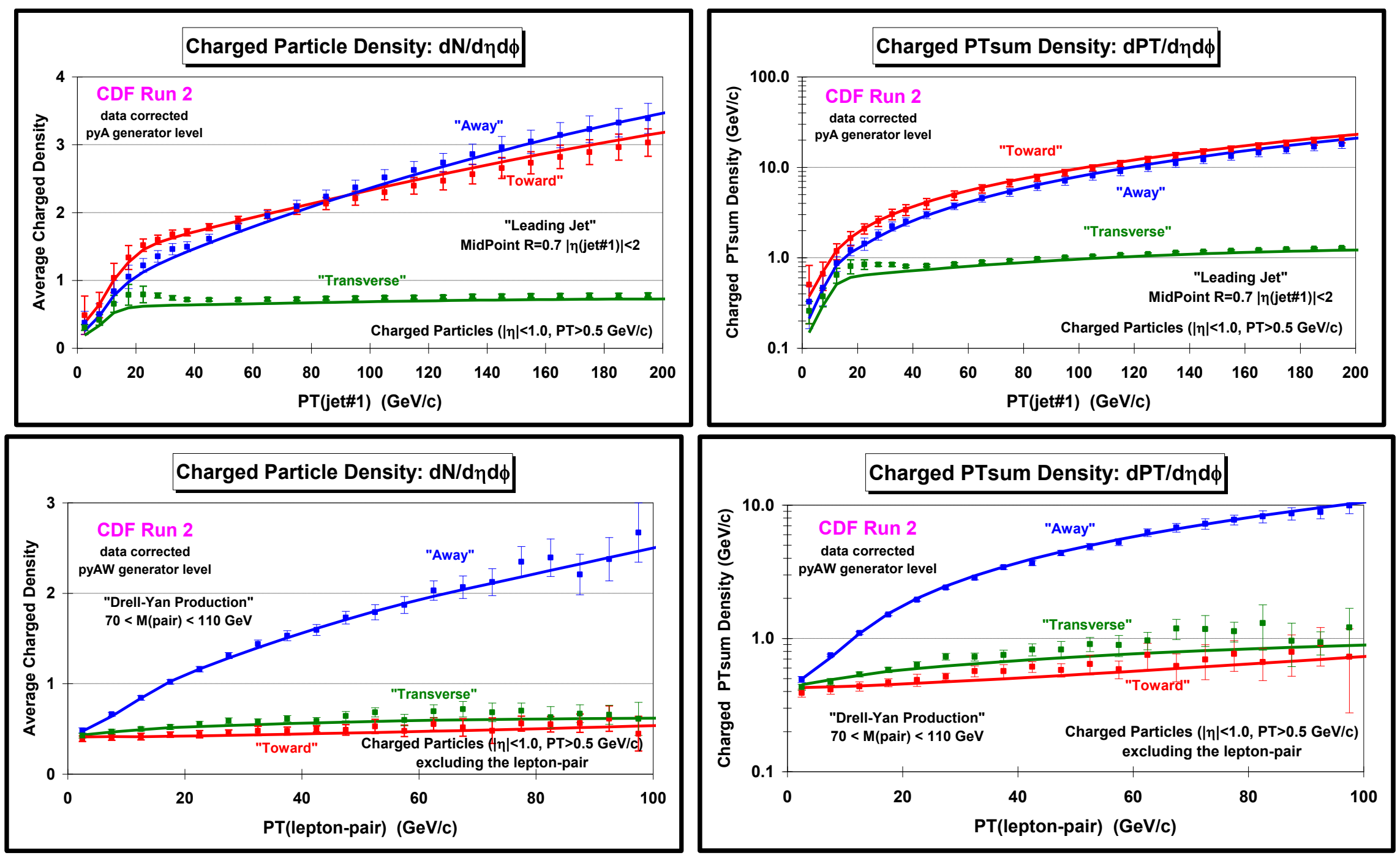

**Fig. 4.1**. CDF data at 1.96 TeV on the density of charged particles, dN/dηdφ, and the scalar PTsum density of charged particles, dPT/dηdφ, with $p_T$ > 0.5 GeV/c and |η| < 1 for leading-jet and Drell-Yan events as a function of the leading-jet $p_T$ and $p_T$(lepton-pair), respectively, for the toward, away, and transverse regions. The data are corrected to the particle level and are compared with PYTHIA Tune A (pyA) and Tune AW (pyAW), respectively, at the particle level.

Figure 4.2 compares the data for leading-jet events with the data for Drell-Yan events for the density of charged particles and the scalar PTsum density, respectively, in the transverse region. The data are compared with PYTHIA Tune A (leading-jet) , Tune AW (Drell-Yan), and HERWIG (without MPI). For large $p_T$(jet#1) the transverse densities are similar for leading-jet and Drell-Yan events as one would expect. HERWIG (without MPI) does not produce enough activity in the transverse region for either process. HERWIG (without MPI) disagrees more with the transverse region of Drell-Yan events than it does with the leading-jet events. This is because there is no final-state radiation in Drell-Yan production so that the lack of MPI becomes more evident.

Fiure 4.3 compares the data for leading-jet events with the data for Drell-Yan events for the average charged-particle $p_T$ and the average maximum charged-particle $p_T$, respectively, in the transverse region. The data are compared with PYTHIA Tune A (leading-jet) , Tune AW (Drell-Yan), and HERWIG (without MPI). MPI provides a hard component to the underlying event and for HERWIG (without MPI) the $p_T$ distributions in the transverse region for both processes are too soft, resulting in an average $p_T$ and average $PT_{max}$ that are too small.

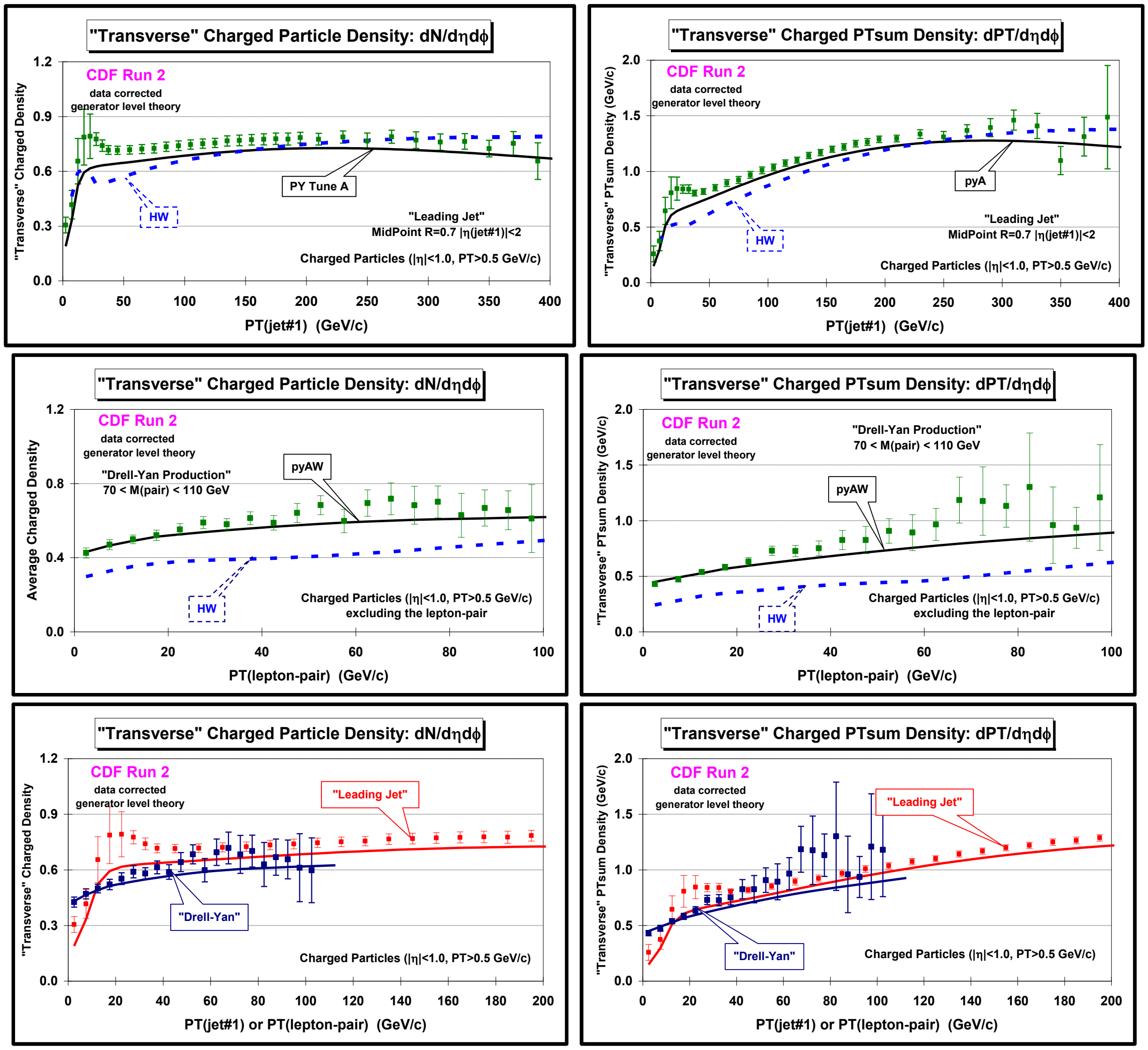

**Fig. 4.2**. CDF data at 1.96 TeV on the density of charged particles, dN/dηdϕ, and the scalar PTsum density of charged particles, dPT/dηdϕ, with $p_T > 0.5$ GeV/c and $|\eta| < 1$ for leading-jet and Drell-Yan events as a function of the leading-jet $p_T$ and $p_T$(lepton-pair), respectively, for the transverse region. The Drell-Yan data are compared with PYTHIA Tune AW (pyAW) and the leading-jet data are compared with PYTHIA Tune A (pyA). Also shown are some prediction from HERWIG without MPI (HW).

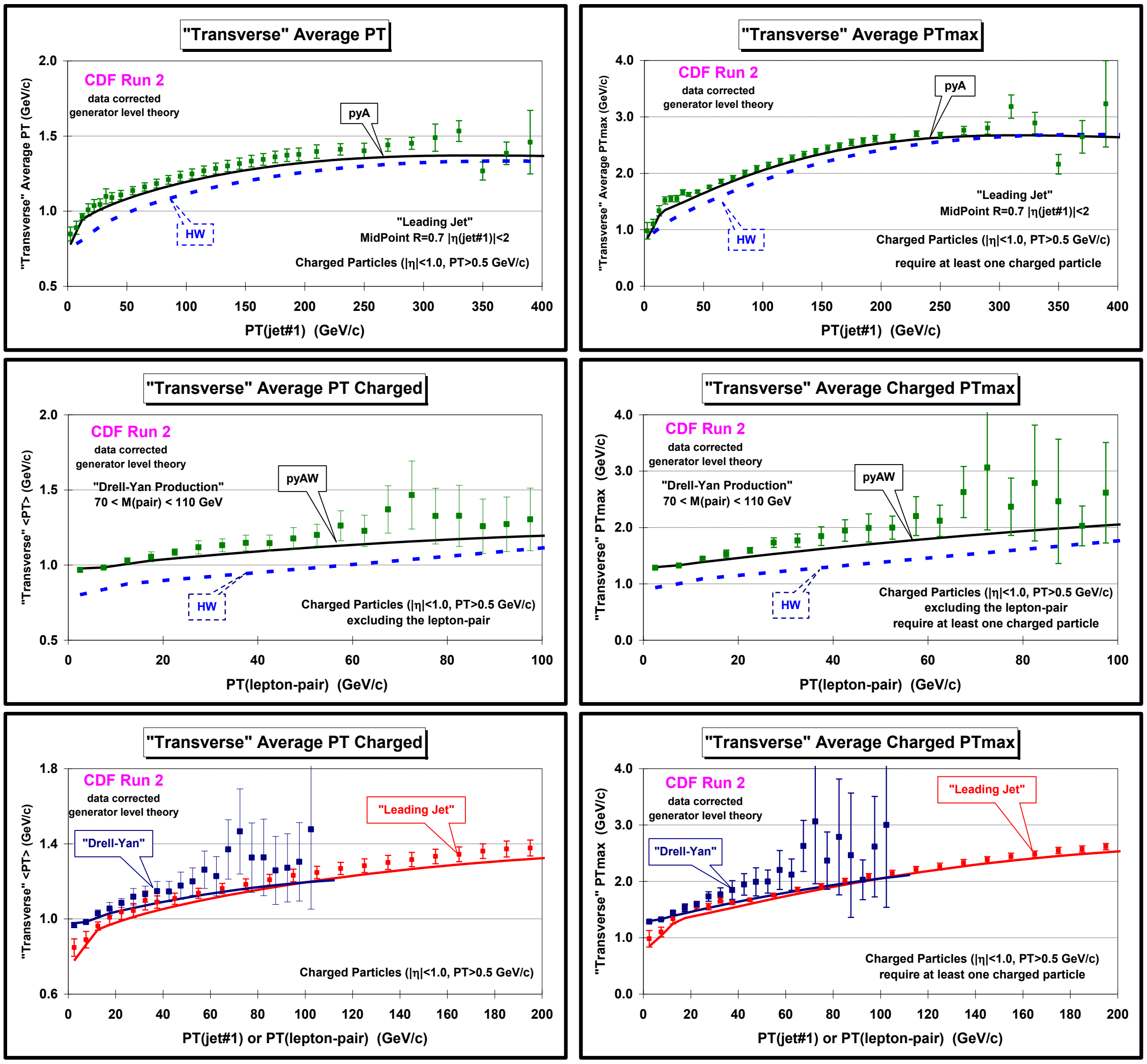

**Fig. 4.3**. CDF data at 1.96 TeV on the average charged particle transverse momentum, $\langle p_T \rangle$, and the average maximum $p_T$, with $p_T > 0.5$ GeV/c and $|\eta| < 1$ for leading-jet and Drell-Yan events as a function of the leading-jet $p_T$ and $p_T$(lepton-pair), respectively, for the transverse region. The Drell-Yan data are compared with PYTHIA Tune AW (pyAW) and the leading-jet data are compared with PYTHIA Tune A (pyA). Also shown are some prediction from HERWIG without MPI (HW).

Figure 4.4 compares the data for leading-jet events with the data for Drell-Yan events for the density of charged particles and the scalar PTsum density, respectively, for the transMAX and transMIN regions. The data are compared with PYTHIA Tune A (leading-jet) , Tune AW (Drell-Yan), and HERWIG (without MPI). For events with large initial-state or final-state radiation the transMAX region would contain the third jet in high-$p_T$ jet production or the second jet in Drell-Yan production. Thus, the transMIN region is very sensitive to the modeling of the underlying event.

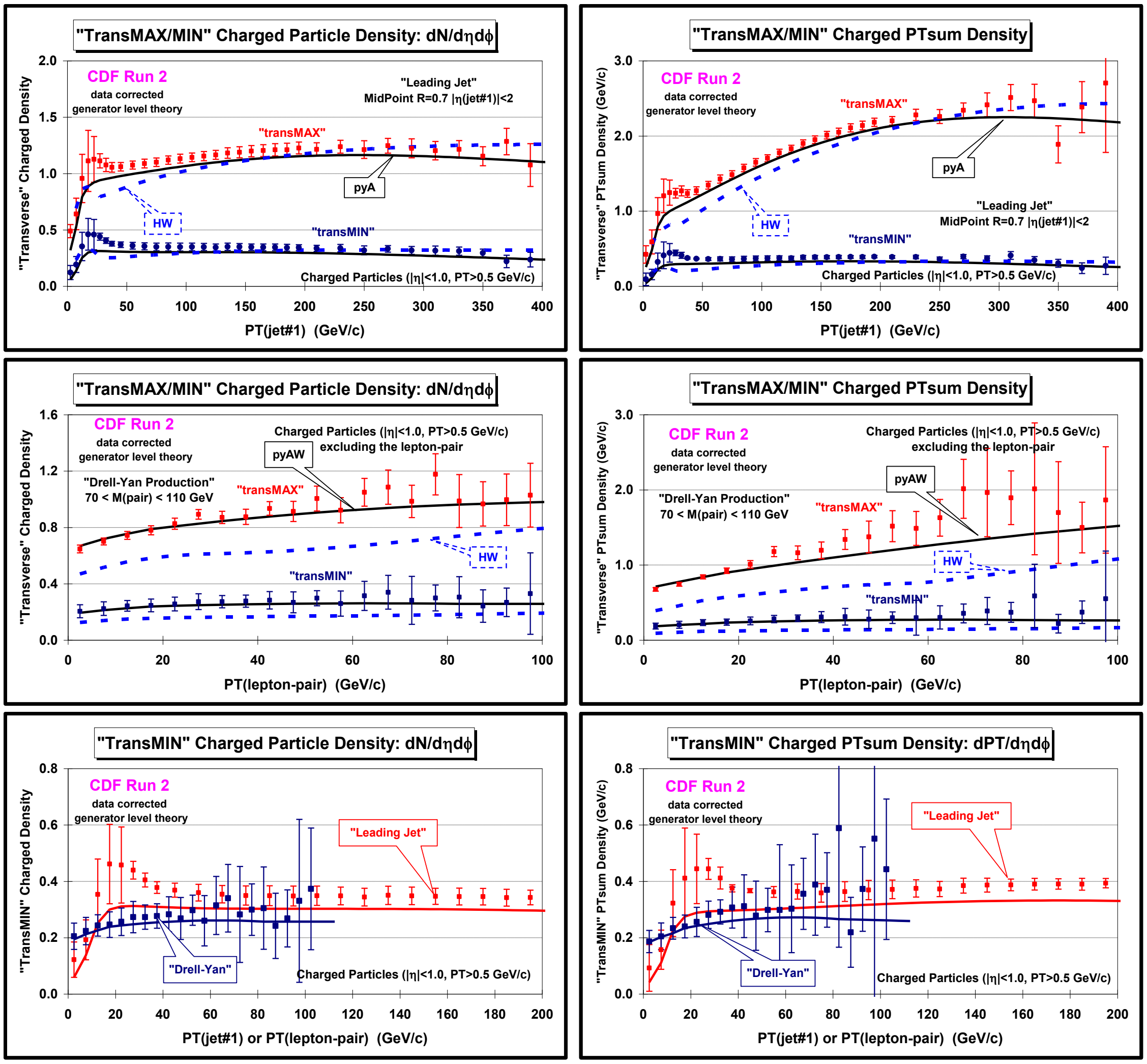

**Fig. 4.4**. CDF data at 1.96 TeV on the density of charged particles, dN/dηdϕ, and the scalar PTsum density of charged particles, dPT/dηdϕ, with $p_T > 0.5$ GeV/c and $|\eta| < 1$ for leading-jet and Drell-Yan events as a function of the leading-jet $p_T$ and $p_T$(lepton-pair), respectively, for the transMAX and transMIN regions. The Drell-Yan data are compared with PYTHIA Tune AW (pyAW) and the leading-jet data are compared with PYTHIA Tune A (pyA). Also shown are some prediction from HERWIG without MPI (HW).

Figure 4.5 compares the data for leading-jet events with the data for Drell-Yan events for the density of charged particles and the scalar PTsum density for transDIF = transMAX - transMIN. The data are compared with PYTHIA Tune A (leading-jet) and Tune AW (Drell-Yan). The transDIF region is sensitive to the hard initial and final-state radiation and is predicted to be very similar in the two processes. Fig. 4.5 also compares the data for leading-jet events with the data for Drell-Yan events for the density of charged particles and the scalar PTsum density in the away region. The away-side jet pseudorapidity distribution and type (quark or gluon) is different for leading-jet and Drell-Yan events so we do not expect the away region to be the same and it is not. However, PYTHIA Tune A and Tune AW describe the data very well.

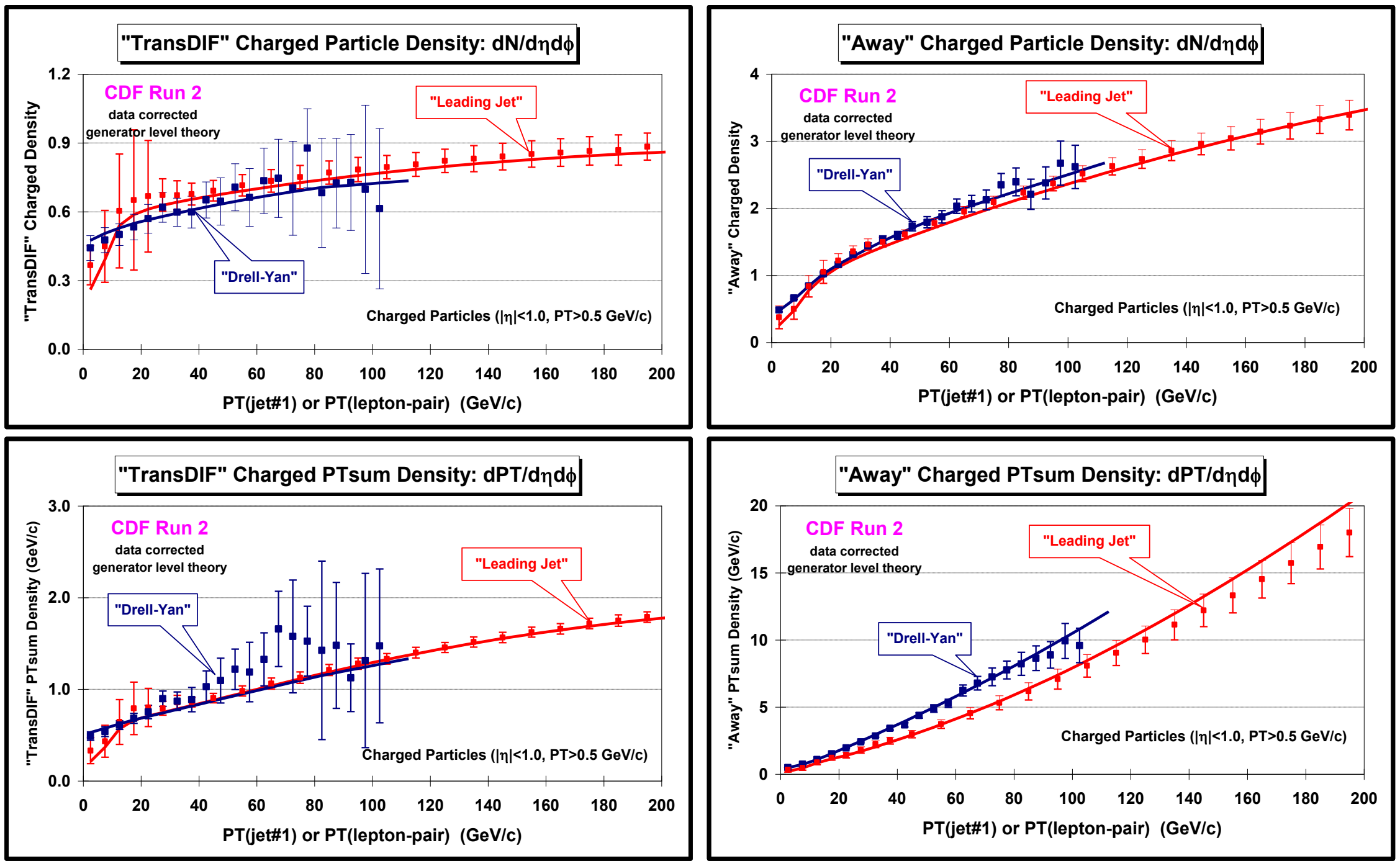

**Fig. 4.5**. CDF data at 1.96 TeV on the density of charged particles, dN/dηdϕ, and the scalar PTsum density of charged particles, dPT/dηdϕ, with $p_T > 0.5$ GeV/c and $|\eta| < 1$ for leading-jet and Drell-Yan events as a function of the leading-jet $p_T$ and $p_T$(lepton-pair), respectively, for the transDIF region (transDIF = transMAX – transMIN) and the away region. The Drell-Yan data are compared with PYTHIA Tune AW (pyAW) and the leading-jet data are compared with PYTHIA Tune A (pyA).

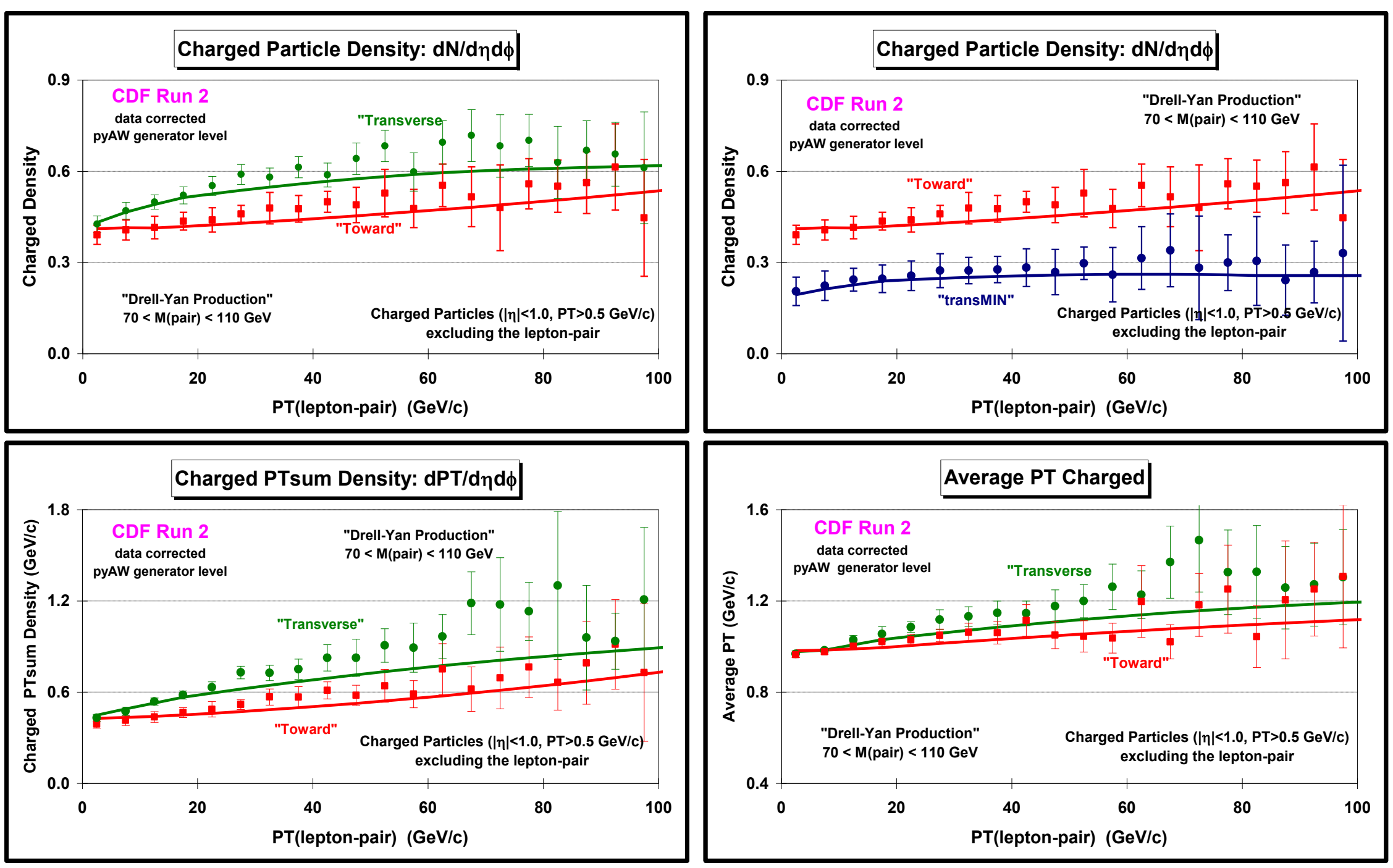

**Fig. 4.6**. CDF data at 1.96 TeV on the density of charged particles, dN/dηdϕ, the scalar PTsum density of charged particles, dPT/dηdϕ, and the average charged-particle $p_T$, with $p_T > 0.5$ GeV/c and $|\eta| < 1$ for Drell-Yan events as a function of $p_T$(lepton-pair) for the toward, transverse, and transMIN regions compared with PYTHIA Tune AW (pyAW).

## (2) The Underlying Event in Drell-Yan Production

Figure 4.6 compares the data in the toward region with the data in the transverse region for Drell-Yan events for the density of charged particles, the scalar PTsum density, and the average charged-particle $p_T$. The data are compared with PYTHIA Tune AW. For high transverse momentum lepton-pair production, particles from initial-state radiation are more likely to populate the transverse region than the toward region and hence the densities are slightly larger in the transverse region. PYTHIA Tune AW describes this very nicely.

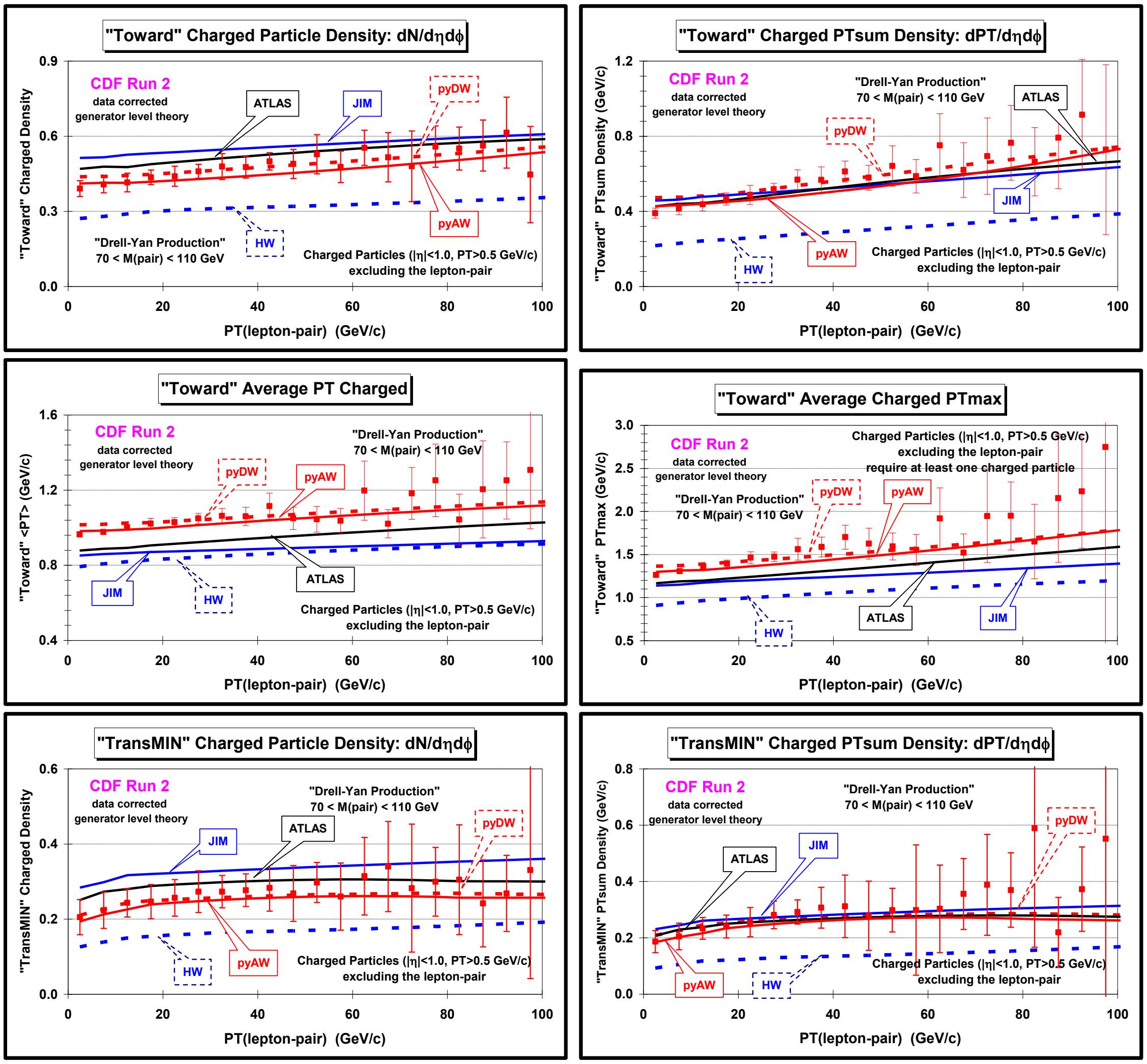

**Fig. 4.7**. CDF data at 1.96 TeV on the density of charged particles, dN/dηdφ, the scalar PTsum density of charged particles, dPT/dηdφ, and the average charged-particle $p_T$, and the average maximum charged-particle $p_T$. with $p_T > 0.5$ GeV/c and $|\eta| < 1$ for Drell-Yan events as a function of $p_T$(lepton-pair) for the toward and the transMIN regions. The data are compared with HERWIG without MPI (HW), HERWIG with JIMMY MPI (JIM), and three PYTHIA Tunes (pyAW, pyDW, ATLAS).

The most sensitive regions to the underlying event in Drell-Yan production are the toward and the transMIN regions, since these regions are less likely to receive contributions from the away-side jet and from initial-state radiation. Fig. 4.7 shows the data for Drell-Yan events for the density of charged particles and the scalar PTsum density, respectively, in the toward and transMIN regions. The data are compared with PYTHIA Tune AW, Tune DW, the PYTHIA ATLAS tune, HERWIG (without MPI), and HERWIG (with JIMMY MPI). The densities are smaller in the transMIN region than in the toward region and this is described well by PYTHIA Tune AW. Comparing HERWIG (without MPI) with HERWIG (with JIMMY MPI) clearly shows the importance of MPI in these regions. Tune AW and Tune DW are very similar. The ATLAS tune and HERWIG (with JIMMY MPI) agree with Tune AW for the scalar PTsum density in the toward and transMIN regions. However, both the ATLAS tune and HERWIG (with JIMMY MPI) produce too much charged-particle density in these regions. The ATLAS tune and HERWIG (with JIMMY MPI) fit the PTsum density, but they do so by producing too many charged particles. They both have too soft a $p_T$ spectrum in these regions. This can be seen clearly in Fig. 4.7 which shows the data for Drell-Yan events on the average charged-particle $p_T$ and the average maximum charged-particle $p_T$, in the toward region compared with the QCD Monte-Carlo models.

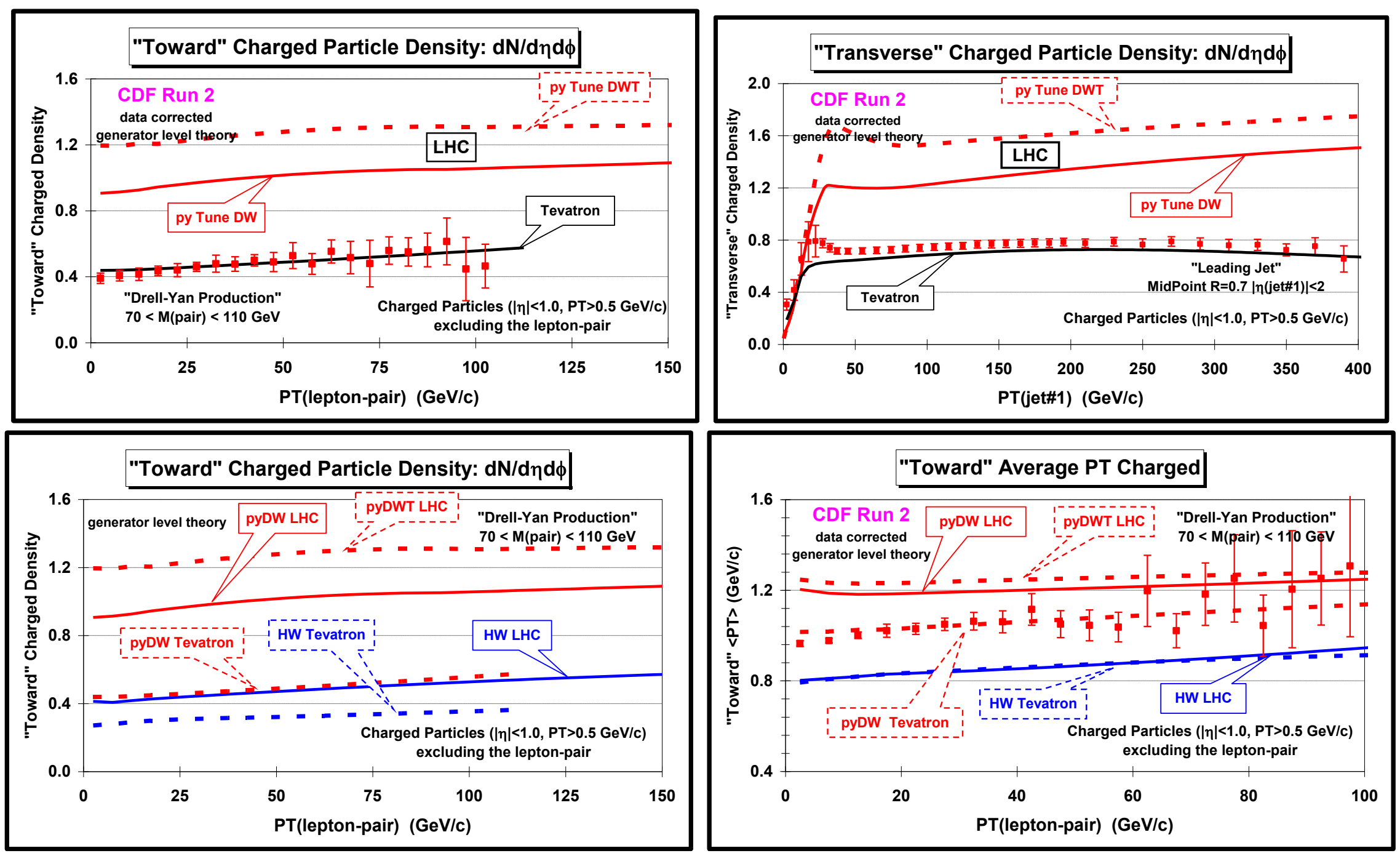

**Fig. 4.8.** CDF data on the density of charged particles, dN/dηdϕ, with $p_T > 0.5$ GeV/c and $|\eta| < 1$ for Drell-Yan events in the toward region and for leading-jet events in the transverse region at 1.96 TeV (Tevatron) together with the predictions of PYTHIA Tune DW (pyDW) and Tune DWT (pyDWT) at the 14 TeV (LHC). Tune DW and Tune DWT are identical at 1.96 TeV (Tevatron). Also shown are CDF data at 1.96 TeV (Tevatron) on the average charged particle transverse momentum, $\langle p_T \rangle$, for Drell-Yan events in the toward region compared with HERWIG without MPI (HW) and PYTHIA Tune DW (pyDW) together with extrapolations to the 14 TeV (LHC).

## (3) Extrapolating to the LHC

Figure 4.8 shows the extrapolation of PYTHIA Tune DW, Tune DWT, and HERWIG (without MPI) to 14 TeV (LHC) for the density of charged particles and the average transverse momentum of charged particles with $p_T > 0.5$ GeV/c and $|\eta| < 1$ in the towards region of Drell-Yan production. The underlying event activity is the same for proton-proton and proton-antiproton collisions. For HERWIG (without MPI) the toward region of Drell-Yan production does not change much in going from the Tevatron to the LHC. Fig. 4.8 also shows the extrapolation of PYTHIA Tune DW and Tune DWT to 14 TeV (LHC) for the transverse density of charged particles ($p_T > 0.5$ GeV/c, $|\eta| < 1$) for leading-jet events. Models with multiple-parton interactions like PYTHIA Tune DW and Tune DWT predict that the underlying event will become much more active (with larger $<p_T>$) at the LHC. PYTHIA Tune DW predicts about a factor of two increase in the activity of the underlying event as measured by the charged-particle density ($p_T > 0.5$ GeV/c, $|\eta| < 1$) in the towards region of Drell-Yan production and the transverse region in leading-jet events. Tune DWT used the default value for PARP(90) and predicts an even greater increase in the activity of the underlying event at the LHC. However, Tune DWT produces less activity than Tune DW in the underlying event at energies below 1.96 TeV and the CDF data at 630 GeV [9] favor Tune DW over Tune DWT.

## (4) $<p_T>$ versus the Multiplicity: Min-Bias and Drell-Yan Events

The total proton-antiproton cross section is the sum of the elastic and inelastic components, $\sigma_{tot} = \sigma_{EL} + \sigma_{IN}$. The inelastic cross section consists of three terms; single diffraction, double-diffraction, and everything else (referred to as the hard core), $\sigma_{IN} = \sigma_{SD} + \sigma_{DD} + \sigma_{HC}$. For elastic scattering neither of the beam particles breaks apart. For single and double diffraction one or both of the beam particles are excited into a high mass color singlet state (*i.e.* $N^*$ states) which then decays. Single and double diffraction also corresponds to color singlet exchange between the beam hadrons. When color is exchanged, the outgoing remnants are no longer color singlets and one has a separation of color resulting in a multitude of quark-antiquark pairs being pulled out of the vacuum. The hard core component, $\sigma_{HC}$, involves color exchange and the separation of color. However, the hard core contribution has both a soft and hard component. Most of the time the color exchange between partons in the beam hadrons occurs through a soft interaction with no high transverse momentum and the two beam hadrons ooze through each other producing lots of soft particles with a uniform distribution in rapidity and many particles flying down the beam pipe. Occasionally there is a hard scattering among the constituent partons producing outgoing particles and jets with high transverse momentum.

Minimum bias (min-bias) is a generic term which refers to events that are selected with a loose trigger that accepts a large fraction of the inelastic cross section. All triggers produce some bias and the term min-bias is meaningless until one specifies the precise trigger used to collect the data. The CDF min-bias trigger consists of requiring at least one charged particle in the forward region $3.2 < \eta < 5.9$ and simultaneously at least one charged particle in the backward region $-5.9 < \eta < -3.2$. Monte-Carlo studies show that the CDF min-bias trigger collects most of the $\sigma_{HC}$ contribution plus small amounts of single and double diffraction [20].

Minimum bias collisions are a mixture of hard processes (perturbative QCD) and soft processes (non-perturbative QCD) and are, hence, very difficult to simulate. Min-bias collisions contain soft beam-beam remnants, hard QCD 2-to-2 parton-parton scattering, and multiple parton interactions (soft & hard). To correctly simulate min-bias collisions one must have the correct mixture of hard and soft processes together with a good model of the multiple-parton interactions. We have seen that multiple parton interactions are a significant component of the underlying event in high $p_T$ jet production and in Drell-Yan lepton-pair production. Multiple-parton interactions are also an important component in min-bias collisions. Min-bias collisions are not the same as the underlying event in a hard-scattering process, since the rate at which MPI occurs is different, but they are related. In selecting a hard-scattering process such as high $p_T$ jet production or in selecting lepton-pair in the mass region of the Z-boson corresponds to selecting a small fraction of min-bias collisions that are very central; the initial proton and antiproton collide with small impact parameter. For these central collisions the probability of additional parton-parton collisions is higher than it is for an average min-bias event.

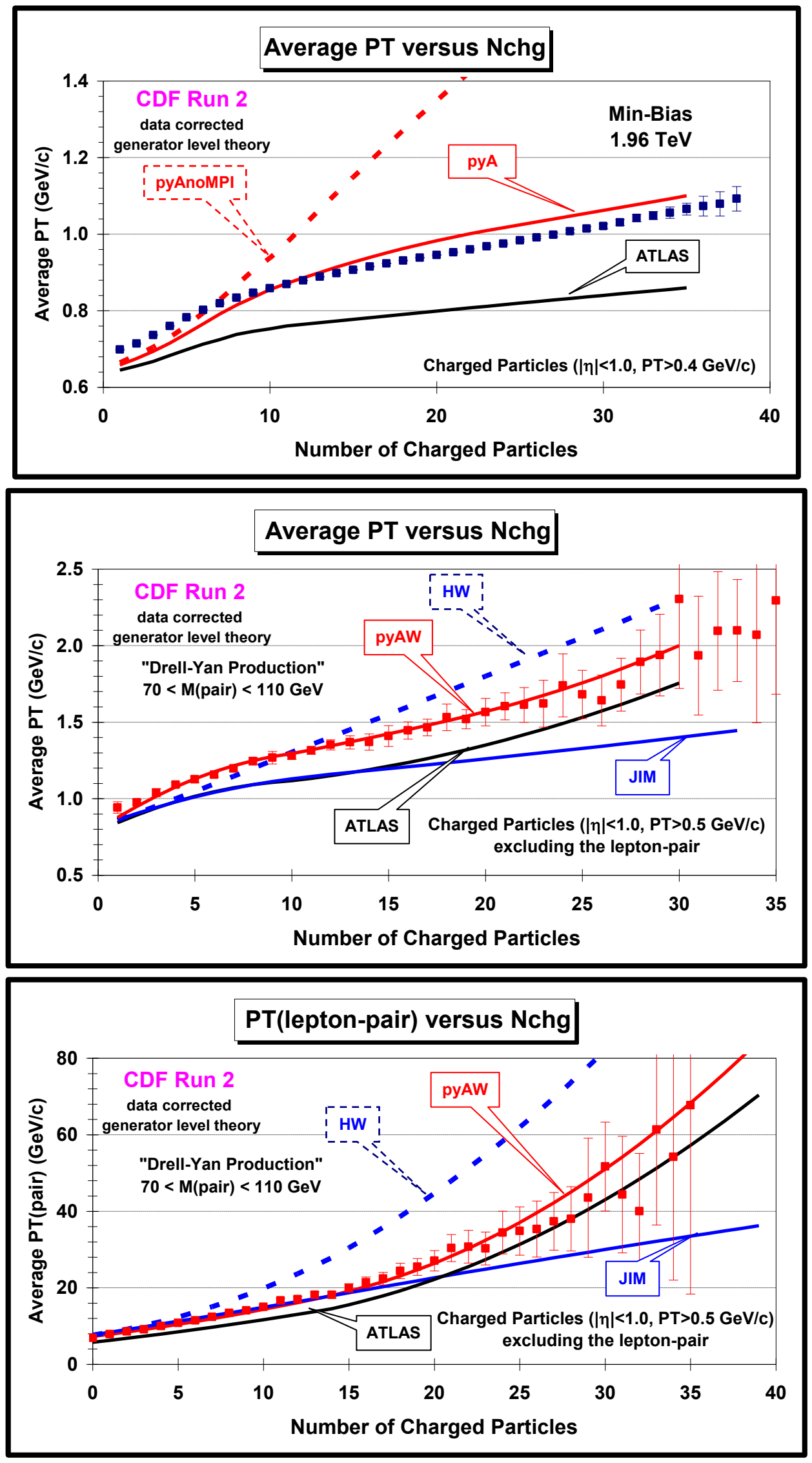

**Fig. 4.9**. (*top*) CDF Min-Bias data on the average $p_T$ of charged particles versus the multiplicity for charged particles with $p_T >$ 0.4 GeV/c and $|\eta| < 1$ from Ref. [20]. The data are compared with PYTHIA Tune A (pyA), the PYTHIA ATLAS tune, and PYTHIA Tune A without MPI (pyAnoMPI). (*middle*) CDF data on the average $p_T$ of charged particles versus the multiplicity for charged particles with $p_T > 0.5$ GeV/c and $|\eta| < 1$ for Drell-Yan events. (*bottom*) CDF data on the average $p_T$ of the lepton-pair versus the multiplicity for charged particles with $p_T > 0.5$ GeV/c and $|\eta| < 1$. The Drell-Yan data are compared with PYTHIA Tune AW, the PYTHIA ATLAS tune, HERWIG without MPI (HW), and HERWIG with JIMMY MPI (JIM).

The first model that roughly described min-bias collisions at CDF was PYTHIA Tune A. However, Tune A was not tuned to fit min-bias collisions. It was tuned to fit the activity in the underlying event in high transverse momentum jet production [4]. However, PYTHIA uses the same $p_T$ cut-off for the primary hard 2-to-2 parton-parton scattering and for additional multiple parton interactions (MPI). Hence, fixing the amount of multiple parton interactions by setting the $p_T$ cut-off allows one to run the hard 2-to-2 parton-parton scattering all the way down to $p_T$(hard) = 0 without hitting a divergence. For PYTHIA the amount of hard scattering in min-bias is, therefore, related to the activity of the underlying event in hard-scattering processes. Neither HERWIG (without MPI) or HERWIG (with JIMMY MPI) can be used to describe min-bias events since they diverge as $p_T$(hard) goes to zero.

Figure 4.9 shows CDF min-bias data corrected to the particle level at 1.96 TeV on the average $p_T$ of charged particles, $<p_T>$, versus the multiplicity for charged particles with $p_T > 0.4$ GeV/c and $|\eta| < 1$ from Ref. [20]. The data are compared with PYTHIA Tune A, the PYTHIA ATLAS tune, and PYTHIA Tune A without MPI (pyAnoMPI). The average $p_T$ is an important observable. The rate of change of $<p_T>$ versus charged multiplicity is a measure of the amount of hard versus soft processes contributing to min-bias collisions and it is sensitive the modeling of the multiple-parton interactions [21]. If only the soft beam-beam remnants contributed to min-bias collisions then $<p_T>$ would not depend on charged multiplicity. If one has two processes contributing, one soft (beam-beam remnants) and one hard (hard 2-to-2 parton-parton scattering), then demanding large multiplicity will preferentially select the hard process and lead to a high $<p_T>$. However, we see that with only these two processes $<p_T>$ increases much too rapidly as a function of multiplicity (see pyAnoMPI). Multiple-parton interactions provide another mechanism for producing large multiplicities that are harder than the beam-beam remnants, but not as hard as the primary 2-to-2 hard scattering. PYTHIA Tune A gives a fairly good description of the $<p_T>$ versus multiplicity, although not perfect. PYTHIA Tune A does a better job describing the data than the ATLAS tune. Both Tune A and the ATLAS tune include multiple-parton interactions, but with different choices for the color connections [22].

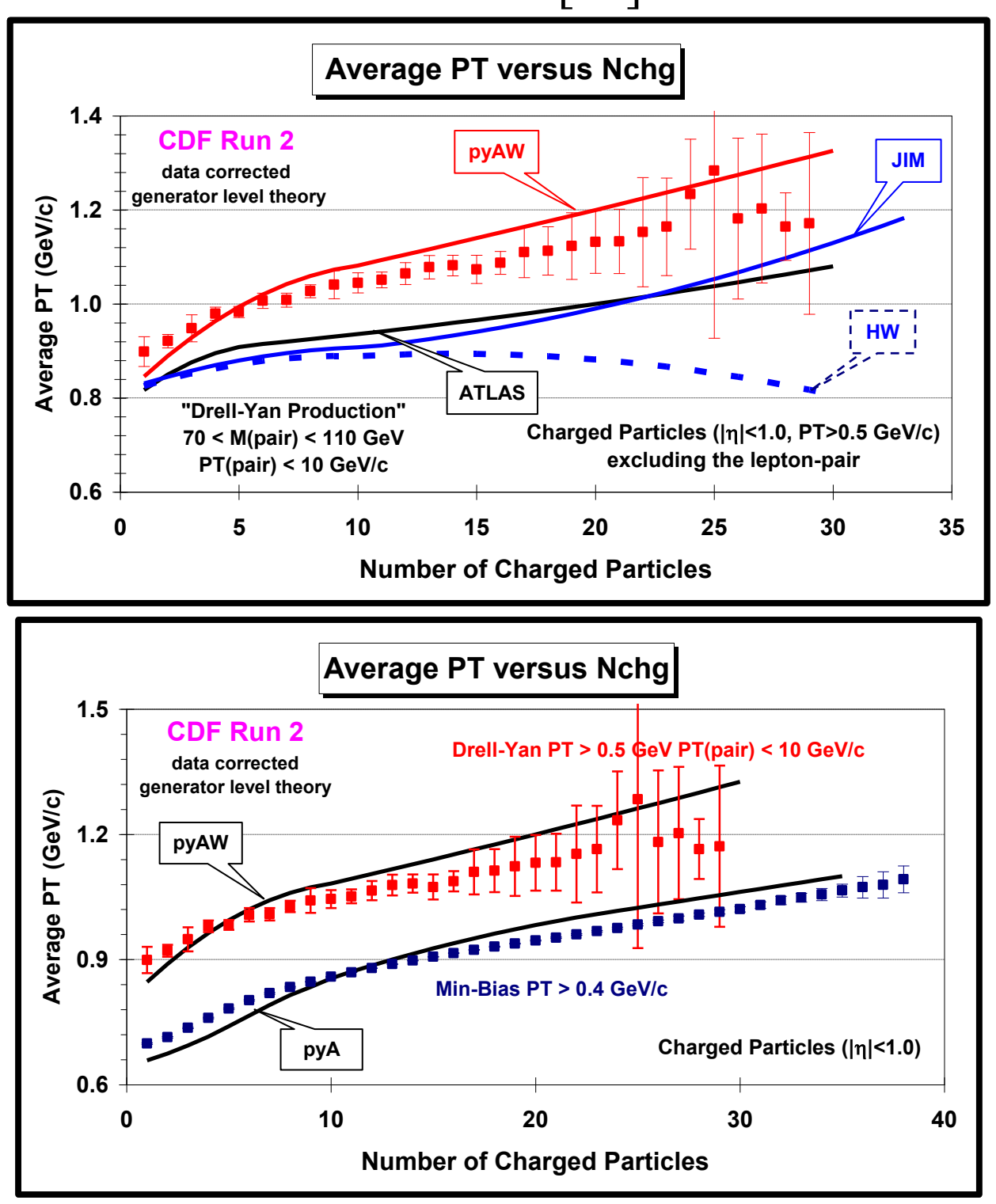

**Fig. 4.10**. (*top*) CDF data at 1.96 TeV on the average $p_T$ of charged particles versus the multiplicity for charged particles with $p_T > 0.5$ GeV/c and $|\eta| < 1$ for Drell-Yan events in which $p_T$(lepton-pair) < 10 GeV/c. The data are compared with PYTHIA Tune AW (pyAW), the PYTHIA ATLAS tune, HERWIG without MPI (HW), and HERWIG with JIMMY MPI (JIM). *(bottom)* Comparison of the average $p_T$ of charged particles versus the charged multiplicity for Min-Bias events from Ref. 20 with the Drell-Yan events with $p_T$(lepton-pair) < 10 GeV/c from this analysis. The Min-Bias data require $p_T > 0.4$ GeV/c and are compared with PYTHIA Tune A (pyA), while the Drell-Yan data require $p_T > 0.5$ GeV/c and are compared with PYTHIA Tune AW (pyAW).

Figure 4.9 also shows the data at 1.96 TeV on the average $p_T$ of charged particles versus the multiplicity for charged particles with $p_T > 0.5$ GeV/c and $|\eta| < 1$ for Drell-Yan events from this analysis. HERWIG (without MPI) predicts the $<p_T>$ to rise too rapidly as the multiplicity increases. This is similar to the pyAnoMPI behavior in min-bias collisions. For HERWIG (without MPI) large multiplicities come from events with a high $p_T$ lepton-pair and hence a large $p_T$ away-side jet. This can be seen clearly in Fig. 4.9 which also shows the average $p_T$ of the lepton-pair versus the charged multiplicity. Without MPI the only way of getting large multiplicity is with high-$p_T$(lepton-pair) events. For the models with MPI one can get large multiplicity either from high-$p_T$(lepton-pair) events or from MPI and hence $<P_T$(lepton-pair)$>$ does not rise as sharply with multiplicity in accord with the data. PYTHIA Tune AW describes the Drell-Yan data fairly well.

Figure 4.10 shows the data at 1.96 TeV on the average $p_T$ of charged particles versus the multiplicity for charged particles with $p_T > 0.5$ GeV/c and $|\eta| < 1$ for Drell-Yan events in which $p_T$(lepton-pair) $< 10$ GeV/c. We see that $<p_T>$ still increases as the multiplicity increases although not as fast. If we require $p_T$(lepton-pair) $< 10$ GeV/c, then HERWIG (without MPI) predicts that the $<p_T>$ decreases slightly as the multiplicity increases. This is because without MPI and without the high $p_T$ away-side jet which is suppressed by requiring low $p_T$ of the lepton pair, large multiplicities come from events with a lot of initial-state radiation and the particles coming from initial-state radiation are soft. PYTHIA Tune AW describes the behavior of $<p_T>$ versus the multiplicity fairly well even when we select $p_T$(lepton-pair) $< 10$ GeV/c.

Figure 4.10 also shows a comparison of the average $p_T$ of charged particles versus the charged multiplicity for min-bias events from Ref. 20 with the Drell-Yan events with $p_T$(lepton-pair) $< 10$ GeV/c. There is no reason for the min-bias data to agree with the Drell-Yan events with $p_T$(lepton-pair) $< 10$ GeV/c. However, they are remarkably similar and described fairly well by PYTHIA Tune A and Tune AW, respectively. This strongly suggests that MPI are playing an important role in both these processes.

# V. Summary & Conclusions

Observables that are sensitive to the underlying event in high transverse momentum jet production (leading-jet events) and Drell-Yan lepton pair production in the mass region of the Z-boson (Drell-Yan events) have been presented and compared with several QCD Monte-Carlo model tunes. The data are corrected to the particle level and compared with the Monte-Carlo models at the particle level. The underlying event is similar for leading-jet and Drell-Yan events as one would expect. This analysis provides data that can be used to test and improve the QCD Monte-Carlo models of the underlying event that are used to simulate hadron-hadron collisions. The data presented here are also important for tuning the new QCD Monte-Carlo multiple-parton interaction (MPI) models [21, 20].

PYTHIA Tune A and Tune AW do a good job in describing the data on the underlying event observables for leading-jet and Drell-Yan events, respectively, although the agreement between theory and data is not perfect. The leading-jet data show slightly more activity in the underlying event than PYTHIA Tune A. PYTHIA Tune AW is essentially identical to Tune A for leading-jet events. All the tunes MPI agree better than HERWIG without MPI. This is especially true in the toward region in Drell-Yan production. Adding JIMMY MPI to HERWIG greatly improves the agreement with data, but HERWIG with JIMMY MPI produces a charged-particle $p_T$ spectrum that is considerably softer than the data. The PYTHIA ATLAS tune also produces a charged-particle $p_T$ spectra that is considerably softer than the data.

The behavior of the average charged-particle $p_T$ versus the charged-particle multiplicity is important. The rate of change of $<p_T>$ versus charged multiplicity is a measure of the amount of hard versus soft processes contributing, and it is sensitive the modeling of the multiple-parton interactions. PYTHIA Tune A and Tune AW do a good job in describing the data on <$p_T$> versus multiplicity for min-bias and Drell-Yan events, respectively, although again the agreement between theory and data is not perfect. The behavior of <$p_T$> versus multiplicity is remarkably similar for min-bias events and Drell-Yan events with $p_T$(lepton-pair) < 10 GeV/c, suggesting that MPI are playing an important role in both these processes.

Models with multiple-parton interactions like PYTHIA Tune DW predict that the underlying event will become much more active (with larger <$p_T$>) at the LHC. For HERWIG (without MPI) the toward region of Drell-Yan production does not change much in going from the Tevatron to the LHC. PYTHIA Tune DW predicts about a factor of two increase in the activity of the underlying event in going from the Tevatron to the LHC as measured by the charged-particle density ($p_T > 0.5$ GeV/c, $|\eta| < 1$) in the towards region of Drell-Yan production and the transverse region in leading-jet events. Tune DWT predicts an even greater increase in the activity of the underlying event at the LHC. However, Tune DWT produces less activity than Tune DW in the underlying event at energies below 1.96 TeV. Tune DW does a better job in fitting the CDF underlying event data at 630 GeV [9], and is hence favored over Tune DWT. At present, PYTHIA tunes with PARP(90) around the value of Tune AW and Tune DW ($\approx 0.25$) seem to be preferred. We will learn a lot about the energy dependence of MPI by comparing the Tevatron results with the early LHC measurements and precise measurements at the LHC require good modeling of the underlying event.

## Acknowledgements

We thank the Fermilab staff and the technical staffs of the participating institutions for their vital contributions. This work was supported by the U.S. Department of Energy and National Science Foundation; the Italian Istituto Nazionale di Fisica Nucleare; the Ministry of Education, Culture, Sports, Science and Technology of Japan; the Natural Sciences and Engineering Research Council of Canada; the National Science Council of the Republic of China; the Swiss National Science Foundation; the A.P. Sloan Foundation; the Bundesministerium für Bildung und Forschung, Germany; the Korean Science and Engineering Foundation and the Korean Research Foundation; the Science and Technology Facilities Council and the Royal Society, UK; the Institut National de Physique Nucleaire et Physique des Particules/CNRS; the Russian Foundation for Basic Research; the Ministerio de Ciencia e Innovación, Spain; the Slovak R&D Agency; and the Academy of Finland.

# References and Footnotes